\documentclass[aps,prd,twocolumn,superscriptaddress]{revtex4-2}
\usepackage{graphicx} 
\usepackage{amsmath, amsthm, amssymb, amsfonts}
\usepackage{nccmath}
\usepackage{graphicx}
\usepackage{float}
\usepackage{array}
\usepackage{verbatim}
\usepackage[margin=1in]{geometry}
\usepackage{color,soul}
\usepackage{fancyhdr}
\usepackage{geometry}
\usepackage{bm}
\usepackage{hyperref}
\hypersetup{colorlinks=true, linkcolor=blue,filecolor=blue,urlcolor=blue, citecolor=blue }
\usepackage[document]{ragged2e}
\usepackage{enumitem}
\usepackage{mathtools}
\usepackage{fancyref}
\usepackage{titlesec}
\usepackage{fix-cm}
\usepackage{cleveref}
\usepackage{esvect}
\usepackage{bbold}
\usepackage{multirow}
\usepackage{array}
\usepackage{braket}
\usepackage{appendix}
\usepackage{comment}
\usepackage{caption}

\begin{document}

\title{Application of the Skyrme Hartree-Fock-Bogoliubov Theory to WIMP-Nucleus Interactions in \textsuperscript{40}Ar}
\author{N. Krishnan}\email{navneet.krishnan@anu.edu.au}
\affiliation{Department of Fundamental and Theoretical Physics, Research School of Physics, Australian National University, ACT, 2601, Australia}
\affiliation{Department of Nuclear Physics and Accelerator Applications, Research School of Physics, Australian National University, ACT, 2601, Australia}
\affiliation{ARC Centre of Excellence for Dark Matter Particle Physics, Australia}
\author{R. Abdel Khaleq}\email{raghda.abdelkhaleq@anu.edu.au}
\affiliation{Department of Fundamental and Theoretical Physics, Research School of Physics, Australian National University, ACT, 2601, Australia}
\affiliation{Department of Nuclear Physics and Accelerator Applications, Research School of Physics, Australian National University, ACT, 2601, Australia}
\affiliation{ARC Centre of Excellence for Dark Matter Particle Physics, Australia}
\author{C. Simenel}\email{cedric.simenel@anu.edu.au}
\affiliation{Department of Fundamental and Theoretical Physics, Research School of Physics, Australian National University, ACT, 2601, Australia}
\affiliation{Department of Nuclear Physics and Accelerator Applications, Research School of Physics, Australian National University, ACT, 2601, Australia}
\affiliation{ARC Centre of Excellence for Dark Matter Particle Physics, Australia}

\date{\today}

\begin{abstract}
\edef\oldrightskip{\the\rightskip}
\justify
\begin{description}
\rightskip\oldrightskip\relax
\setlength{\parskip}{0pt}
WIMP scattering from \textsuperscript{40}Ar is investigated using a self-consistent Skyrme Hartree–Fock–Bogoliubov (HFB) approach. Nuclear form factors relevant to dark matter direct detection are calculated from the resulting one-body density matrix elements and compared with shell-model predictions. Good agreement is found for the spin-independent response, while significant differences are observed for the spin-orbit response due to variations in single-particle occupancies. The effects of particle-number projection are shown to be small for \textsuperscript{40}Ar. These results demonstrate the sensitivity of certain dark matter response channels to the underlying nuclear structure model and establish a framework for extending mean-field calculations to nuclei beyond the reach of large-scale shell-model studies.
\end{description}
\end{abstract}
\maketitle

\tableofcontents

\justify

\section{Introduction}

\noindent The Weakly Interacting Massive Particle (WIMP) remains an attractive candidate for dark matter (DM). While the original supersymmetric realisation of the WIMP has not shown any experimental data confirming its existence \cite{Bertone_Tait_2018}, the WIMP as a general class of models---a high-mass spin-1/2 fermion with interactions with Standard Model (SM) particles below the weak scale---is an ongoing possibility. The search for such a model requires two simultaneous approaches: experiments searching for evidence of WIMP-SM interactions, and theoretical models able to interpret any data or null results and guide future experiments.
\\
\\
In this paper we focus on the direct detection class of experiments, 
which seek to observe the interactions of WIMPs with nuclei in terrestrial detectors. As direct detection experiments search for interactions between WIMPs and nuclei, understanding the effect of nuclear structure on the overall interaction is crucial to ruling out proposed models of dark matter, and to correctly interpreting data in the event of a positive detection.
\\
\\
To date, the nuclear shell model with a phenomenological potential has been the standard approach to incorporating nuclear structure effects in WIMP-nucleus interactions \cite{Fitzpatrick:2012ix,AbdelKhaleq:2023ipt,Hoferichter,Holt}. 
Indeed, the shell-model aims to describe the nuclear wave-function in the laboratory frame, e.g., as seen by an external particle interacting with the nucleus. It is thus \textit{a priori} well-suited to the context of WIMP-nucleus interaction. A limitation of the shell model, however, comes from the assumption of an inert spherical core to reduce the numerical calculations to a limited valence space. 
This approach is particularly suited to nuclei near closed nuclear shells, whose shape tends to be closer to sphericity. However, it falls short when considering nuclei away from magic numbers (such as germanium isotopes relevant to SuperCDMS \cite{SuperCDMS}, EDELWEISS \cite{EDELWEISS} and CDEX \cite{CDEX}) which can be significantly deformed. 
In addition, while $^{128-132}_{54}$Xe  (XENONnT \cite{XENON:2023cxc} and XLZD \cite{XLZD}) and $^{127}_{53}$I isotopes (ANAIS \cite{Amare:2021yyu}, COSINE \cite{COSINE-100:2021zqh} and SABRE \cite{SABRE}) are only a few nucleons away from the doubly magic $^{132}_{50}$Sn nucleus, the valence spaces require important restrictions that may have significant impact on the resulting nuclear form factors \cite{AbdelKhaleq:2024hir}.
\\
\\
To remedy this issue, we instead consider self-consistent mean-field approaches, such as the Hartree-Fock-Bogoliubov (HFB) theory, to compute the nuclear ground-state wave-function and use it in modelling WIMP-nucleus interactions. Unlike the shell model, the HFB theory does not assume a core and instead treats all nucleons on the same footing. However, the wave-function is constrained to a quasi-particle vacuum, thus missing some correlations included in the shell model. Nevertheless, several correlations can be included in mean-field calculations thanks to  symmetry breaking techniques \cite{sheikh2021,cesca2023}. For example, short range correlations responsible for superfluidity in nuclei are accounted for by breaking $U(1)$ gauge invariance. Similarly, long-range correlations inducing nuclear deformations are included by breaking rotational invariance and reflection symmetry for example, leading to quadrupole and octupole deformations respectively. However, while symmetry breaking is a powerful technique to account for such correlations, the price paid to retain the simplicity of a mean-field description is that the nuclear state no longer has good quantum numbers associated with these symmetries. In particular, HFB states of superfluid, quadrupole- and octupole-deformed nuclei are characterised by coherent superpositions of particle numbers, angular momenta, and parities, respectively. In fact, a symmetry broken HFB ground-state provides a representation of the nucleus in its \textit{intrinsic frame}, i.e., as seen by a nucleon within the nucleus, while the considered symmetries are expected to hold in the \textit{lab frame}. This description is in principle incompatible with the problem at hand, where a WIMP is interacting with a nucleus in the lab frame. It is then necessary to perform a change from the intrinsic to the lab frame by restoring the broken symmetries \cite{sheikh2021}. This is achieved using projection techniques that restore the good quantum numbers associated with each symmetry. In practice, particle number, angular momentum and parity are restored by summing quasiparticle {vacua} 
with various gauge angles, orientations and reflection images, respectively. 
\\
\\
Using the code \textsc{hfbtho}  
\cite{HFBTHO4}, we compute the nuclear form factors required to determine the WIMP-nucleus cross section searched for in direct detection. To be confident in applying this approach, we must first understand how its output differs to the shell model. In this work, we thus present a proof-of-principle of this method using the \textsuperscript{40}Ar nucleus as a target, relevant to several direct detection experiments (e.g., DarkSide \cite{DarkSide-50:2022qzh} and DEAP \cite{DEAP:2019yzn}). Indeed, being superfluid, even-even, and spherical, $^{40}$Ar offers an ideal benchmark to compare form factors from mean-field approaches against those predicted with the shell model. 
Applications to deformed nuclei 
will be considered in future works.
\\
\\
This paper is structured as follows. In Section~\ref{sec:formalism} we summarise the DM-nucleus scattering formalism employed here, as derived by Fitzpatrick \textit{et al.} in \cite{Fitzpatrick:2012ix,Anand:2013yka}. In Section~\ref{sec:SM}, we quickly review the  shell model formalism as applied in \cite{AbdelKhaleq:2023ipt,AbdelKhaleq:2024hir} to compute the DM-nucleus form factors. This formalism is applied in this paper to \textsuperscript{40}{Ar} for comparison with the HFB results. In Section \ref{sec:MF} we introduce the HFB approach to computing the nuclear ground-state wave functions, and detail how the formalism typically employed in shell model form factor calculations is modified to compute the corresponding mean-field counterparts. In Section \ref{sec:results} we present and discuss the results of our work before concluding in Sec.~\ref{sec:conclusions}.

\section{Dark Matter-Nucleus Scattering Formalism\label{sec:formalism}}

\noindent Direct detection experiments measure the differential recoil scattering rate, typically in units of counts per day/kg/keV \cite{Undagoitia_Rauch_2015}:
\begin{equation}
    \frac{{\rm d}R}{{\rm d}E_R} = \frac{N_T \rho_\chi}{m_\chi} \int_{v>v_{\text{min}}} v f(\vec{v}) \ \frac{{\rm d}\sigma_T}{{\rm d}E_R} (\vec{v}) \ {\rm d}^3v. \label{eq: standard rate expression}
\end{equation}
\noindent Here, $E_R$ is the nuclear recoil energy, $m_\chi$ is the WIMP mass, $N_T$ is the number of target nuclei of type $T$ per detector mass, ${\rm d}\sigma_T/{\rm d}E_R$ is the differential cross section, $\rho_\chi$ is the  DM density on Earth, and $f(\vec{v})$ is the WIMP halo velocity distribution in the Earth reference frame. $v_{\min}$ is the minimum DM velocity required to induce an elastic nuclear recoil at energy $E_R$:
\begin{equation}
    v_{\text{min}} (E_R) = \frac{q}{2\mu_T} = \frac{1}{\mu_T} \sqrt{\frac{m_T E_R}{2}},
\end{equation}
\noindent where $q$ is the momentum transfer and  $\mu_T= m_T m_\chi/(m_T+m_\chi)$ is the WIMP-nucleus reduced mass ($m_{T}$ is the mass of the target nucleus and $m_\chi$ is the WIMP mass). \\
\\
The kinematics of WIMP-nucleus scattering are described by the momentum transfer $q$ between the WIMP $\chi$ and a given nucleon $N$ in an interaction:
\begin{equation}\label{momentum transfer}
    \vec{q}= \vec{p}\hspace{0.5mm}'-\vec{p}= \vec{k}-\vec{k}',
\end{equation}
\noindent where $\vec{p}\hspace{0.5mm}'$ ($\vec{p}$) is the outgoing (incoming)  momentum of $\chi$ and $\vec{k}'$ ($\vec{k}$) is the outgoing (incoming) momentum of $N$. 
\\
\\
\noindent Given an interaction Hamiltonian $\mathcal{H}_{int}$, the {spin-averaged squared scattering matrix element} 
$\mathcal{M}$ is given by:
\begin{widetext}
\begin{align}
    \mathcal{M} &= \frac{1}{2J_i+1} \sum_{M_i, M_f} \big| \langle J_i M_f  | \sum_{n=1}^A \mathcal{H}_{\text{int}} (\vec{x}_n) |  J_i M_i \rangle \big|^2 \nonumber
    \\
    &= \frac{4\pi}{2J_i+1} \Bigg[ \sum_{\{j, X\}} \sum_{J} |\langle J_i || l_j \ X_J (q)|| J_i \rangle |^2  + \sum_{\substack{\{j, X\}; \{k, Y\}; \\
    \{X, Y\}, X \neq Y}}  \sum_{J} \text{Re} \big[  \langle J_i || l_j \ X_J (q) || J_i \rangle \langle J_i || l_k \ Y_J (q) || J_i \rangle^* \big]\Bigg]. 
     \label{scatamp}
\end{align}
\end{widetext}

\noindent The right hand side of Eq.~(\ref{scatamp}) is decomposed into a sum over operators $X,Y$ characterising the nuclear interaction response. These are derived  from a general non-relativistic effective field theory (EFT) of WIMP-nucleus interactions in Refs.~\cite{Fitzpatrick:2012ix,Anand:2013yka}. The total EFT Lagrangian can be expressed as a sum $\sum\limits_i c_i \mathcal{O}_i$ over {15 possible nucleon and WIMP operators $\mathcal{O}_i$. However, for a nucleus such as \textsuperscript{40}Ar with ground state  spin-parity $J_i^\pi=0^+$, only 7 of these nucleon-level operators are relevant. Six of them can arise from an interaction mediated by a spin $\leq 1$ particle:}
\begin{equation}\label{eq: NR O operators}
\begin{split}
    &  \mathcal{O}_1= \mathbf{1}, \hspace{8mm} \mathcal{O}_3= i \vec{S}_N \cdot \left(\frac{\vec{q}}{{m_N}} \times \vec{v}^\perp \right), \\
    & \mathcal{O}_5=  i \vec{S}_\chi \cdot \left(\frac{\vec{q}}{{m_N}} \times \vec{v}^\perp \right), \hspace{8mm} \mathcal{O}_8=  \vec{S}_\chi \cdot  \vec{v}^\perp, \hspace{8mm}  \\
    &\mathcal{O}_{11} = i \vec{S}_\chi \cdot \frac{\vec{q}}{{m_N}}, \hspace{8mm} \mathcal{O}_{12} = \vec{S}_\chi \cdot \left(\vec{S}_N  \times \vec{v}^\perp \right). \\
\end{split}
\end{equation}

\noindent {One additional operator contributes for a spin $> 1$ mediator: 
\begin{equation}
    \mathcal{O}_{15} = -\left( \vec{S}_{\chi} \cdot \frac{\vec{q}}{m_N}\right) \left[ \left( \vec{S}_N \times \vec{v}^{\perp}\right)\cdot\frac{\vec{q}}{m_N}\right].
\end{equation}
Although this operator is highly suppressed relative to the others, being cubic in the particle velocity, it is included in the Mathematica package \cite{Anand:2013yka} used in this work.} 
\\
\\
\noindent Above, $\vec{S}_{\chi,N}$ denotes the spin of the particles. The perpendicular velocity has the form
\begin{equation}\label{v perp expression}
\begin{split}
    \vec{v}^{\perp} & \equiv \vec{v} +\frac{\vec{q}}{2 \mu_N}\\
    & = \frac{1}{2} \left( \vec{v}_{\chi, \text{in}} +\vec{v}_{\chi, \text{out}} -\vec{v}_{N, \text{in}} -\vec{v}_{N, \text{out}} \right)\\
    & =  \frac{1}{2} \left(\frac{\vec{p}}{m_\chi} +\frac{\vec{p}\hspace{0.5mm} '}{m_\chi} -\frac{\vec{k}}{m_N} -\frac{\vec{k}'}{m_N} \right),
\end{split}
\end{equation}
where $m_N$ is the nucleon mass. A similar perpendicular velocity is defined for the target nucleus, 
\begin{equation}\label{vTperp eqn}
\begin{split}
    \vec{v}_T^\perp & = \frac{1}{2} (\vec{v}_{\chi, \text{in}} + \vec{v}_{\chi, \text{out}} - \vec{v}_{T, \text{in}} - \vec{v}_{T, \text{out}}) \\
    & = \vec{v}_T+\frac{\vec{q}}{2\mu_T},
\end{split}
\end{equation}
where, $\vec{v}_{T, \text{in(out)}}$ = {$\frac{1}{m_T}\sum_{j=1}^A \vec{p}_{N,\text{in(out)}}(j)$}.
\\
\\
While the general Lagrangian would generate 6 nuclear operators $X$ that contribute to the scattering cross section, only two gain contributions from the above nucleon-level operators when the nuclear ground state has $J_i=0$. These are:
\begin{equation}\label{eq: nuclear operators def}
\begin{split}
     M_{JM} (q\vec{x}) & \equiv j_J(qx) Y_{JM} (\Omega_x), \\
    \Phi''_{JM} (q \vec{x}) & \equiv i \left(\frac{\vec{\nabla}}{q} M_{JM} (q\vec{x}) \right)\cdot \left( \vec{\sigma}_N \times \frac{1}{q} \vec{\nabla} \right), 
\end{split}
\end{equation} 
where $\vec{\sigma}_N$ is the nucleon spin operator. The operator $M$ encodes the spin-independent response, which produces a cross section scaling with the atomic mass as $A^2$. The operator $\Phi''$ can similarly be understood at leading order to encode the contribution of the spin-orbit interaction of the nucleons. 
\\
\\
\noindent The coefficients $l$ in Eq.~(\ref{scatamp}) encode the dark matter contributions to the interaction, each associated with a specific nuclear operator $X$. A sum is taken over the nucleons $n$ in the nucleus to determine the total response. The reduced matrix elements in Eq.~(\ref{scatamp}) are computed from the full matrix elements via the Wigner-Eckart theorem:
\begin{equation}
\begin{split}
   & \langle j' m' | T_{JM}  | j m \rangle 
    \\
    &= (-1)^{j'-m'} \begin{pmatrix}
         \vspace{2mm} j'  &  J  &  j \\
         \vspace{2mm} -m' &  M &  m  \hspace{2mm}
         \end{pmatrix} \langle j' || T_J || j \rangle.
\end{split}
\end{equation}


\noindent From the above expression for the scattering amplitude, the differential cross section can be written:
\begin{equation}\label{eq: anand cross section}
     \frac{{\rm d}\sigma_T}{{\rm d}E_R} = \frac{m_T}{2\pi v^2} \sum_{i,j} \sum_{N,N'=p,n} \mathfrak{c}_i^{(N)} \mathfrak{c}_j^{(N')} F_{i,j}^{(N,N')} (v^2,q^2).
\end{equation}
\noindent The Wilson coefficients $\mathfrak{c}$ now encode the dark matter contributions to the interaction, and their expression is informed by the particular WIMP model in question. The form factors $F$ encode the nuclear contribution, with form factors for single-operator responses coming from the first term in Eq.~(\ref{scatamp}), and form factors from two-operator interference generated by the second term. Defining:
\begin{align} \label{Eq:FF}
    F_X^{(N,N')} &= \frac{4\pi}{2J_i+1} \sum\limits_{J=0}^{2J_i} \braket{J_i \lvert\lvert X^{(N)}_J \lvert \lvert J_i} \braket{J_i \lvert\lvert X^{(N')}_J \lvert \lvert J_i},
    \\
    F_{X,Y}^{(N,N')} &= \frac{4\pi}{2J_i+1} \sum\limits_{J=0}^{2J_i} \braket{J_i \lvert\lvert X^{(N)}_J \lvert \lvert J_i} \braket{J_i \lvert\lvert Y^{(N')}_J \lvert \lvert J_i},
\end{align}
\noindent where in this case $X,Y=\{M, \Phi'' \}$, the relevant form factors are then:

\begin{equation}
\label{eq:anan_ff}
    \begin{split}
         & F_{1,1}^{(N,N')}  = F_{M}^{(N,N')}, \hspace{4mm} \\
         & F_{3,3}^{(N,N')}  = \frac{q^4}{4 m_N^4} 
        F_{\Phi''}^{(N,N')}   
        , \\
         & F_{5,5}^{(N,N')} =  C(j_{\chi}) \frac{q^2}{4 m_N^2} {v^\perp_T}^2 F_{M}^{(N,N')} 
        , \hspace{4mm} \\
        & F_{8,8}^{(N,N')}  =  \frac{C(j_{\chi})}{4} {v^\perp_T}^2 F_{M}^{(N,N')} 
        , \\ 
         & F_{11,11}^{(N,N')}  = C(j_{\chi})\frac{q^2}{4 m_N^2} F_{M}^{(N,N')} , \\
        & F_{12,12}^{(N,N')}  = 
        C(j_{\chi})\frac{q^2}{16 m_N^2}  F_{\Phi''}^{(N,N')},
        \\
         & F_{1,3}^{(N,N')}   = \frac{q^2}{2m_N^2}F_{M,\Phi''}^{(N,N')}, \\
         & F_{11,12}^{(N,N')}  = C(j_{\chi}) \frac{q^2}{8 m_N^2} F_{M, \Phi''}^{(N,N')}. \\
    \end{split}
\end{equation}
\noindent Here, $j_\chi$ is the WIMP spin quantum number and $C(j_{\chi})=4j_\chi (j_\chi+1)/3$ is normalised for spin-1/2 particles such that $C(1/2)=1$. $N,N' = \{ p,n \}$ denote the particular nucleon in question. The proton and neutron nuclear operators are given by $X_J^{(p)}= \frac{1+\tau_3}{2} \ X_J$ and $X_J^{(n)}= \frac{1-\tau_3}{2} \ X_J$, where $\tau_3$ is the usual third component of isospin. We have $F^{(p,n)}_{X,Y} (q^2)= F^{(n,p)}_{X,Y} (q^2)$ only for operator responses with $X=Y$.
\\
\\

\noindent Computing the cross section is thus a matter of fixing the coefficients $\mathfrak{c}$ (either inputting direct numerical values or deriving them from a specific high-energy model) and computing the form factors $F$ from a nuclear structure model. The latter is the focus of this work.

\section{The Shell Model Approach\label{sec:SM}}

\noindent Prior to discussing the mean-field formalism used in our work, we provide a summary of the shell model approach \cite{andrew}. This serves as a benchmark to which the mean-field calculations can be compared. Additionally, much of the formalism and code used to compute the form factors in the shell model can be applied to the mean-field case, with some adaptations.
\\
\\
The nuclear matrix elements required to compute the form factors can be expressed in terms of single-nucleon matrix elements and One-Body Density Matrix Elements (OBDMEs) $\Psi^{J;\tau}_{|\alpha|, |\beta|}$:
\begin{widetext}
\begin{equation} \label{reduced nuclear matrix element}
\begin{split}
     \langle J_i; T M_T \ ||  \sum_{m=1}^{A} \hat{O}_{J, \tau} (q \vec{x}_m) ||  \ J_i ; T M_T \rangle  &= (-1)^{T-M_T}  \begin{pmatrix}
         \vspace{2mm} T  &  \tau  &  T \\
         \vspace{2mm} -M_T &  0 &  M_T  \hspace{2mm}
         \end{pmatrix}  
         \langle J_i; T \ \vdots \vdots \sum_{m=1}^{A} \hat{O}_{J, \tau} (q \vec{x}_m) \vdots \vdots \ J_i ; T \rangle 
         \\
  &=    (-1)^{T-M_T}  \begin{pmatrix}
         \vspace{2mm} T  &  \tau  &  T \\
         \vspace{2mm} -M_T &  0 &  M_T  \hspace{2mm}
         \end{pmatrix}
         \sum_{|\alpha|, |\beta|} \Psi^{J;\tau}_{|\alpha|, |\beta|}  \langle |\alpha| \ \vdots \vdots \hat{O}_{J, \tau} (q \vec{x}) \vdots \vdots \ |\beta| \rangle.
\end{split}
\end{equation}
\end{widetext}
Above, $\alpha$ and $\beta$ are single-nucleon states with the usual quantum numbers $\beta = \{n_\beta, l_\beta, j_\beta, m_{j_\beta}, m_{t_\beta} \}$, and the reduced states have quantum numbers $|\beta|= \{n_\beta, l_\beta, j_\beta\}$. The nucleon isospin state $t_\beta=t_\alpha=1/2$ is assumed. The total nuclear isospin is denoted by $T$ and its projection by $M_T$. 
\\
\\
The symbol $\vdots \vdots$ denotes a matrix element reduced in both angular momentum and isospin via the Wigner-Eckart theorem. Additionally, $\tau=\{0, 1\}$ with $ \hat{O}_{J, \tau} =  \hat{O}_{J} \ \tau_3^\tau$. The expression can thus be separated into $\tau=0$ (isospin-independent) and $\tau=1$ (isospin-dependent) terms. 
\\
\\
The OBDMEs are given by:
\begin{equation}
\begin{split}
    \Psi^{J;\tau}_{|\alpha|, |\beta|} & \equiv \frac{\langle J_i; T \ \vdots \vdots \left[a^\dagger_{|\alpha|} \otimes \tilde{a}_{|\beta|} \right]_{J;\tau} \vdots \vdots \ J_i ; T \rangle}{\sqrt{(2J+1)(2\tau+1)}},\label{eq:obdme}
\end{split}
\end{equation}
where $\tilde{a}_{\beta }=(-1)^{j_\beta -m_{j_\beta}+1/2-m_{t_\beta}} \ a _{|\beta|;-m_{j_\beta}, -m_{t_\beta}}$ and $\otimes$ denotes a tensor product. The OBDMEs are of course dependent on a particular nuclear structure model, and will lead to different form factors between the shell model and mean-field approaches. In the shell model case, these OBDMEs are obtained from large-scale shell model calculations using NuShellX \cite{Brown:2014bhl}, a program which evaluates nuclear wave functions in addition to various nuclear observables. This shell model approach assumes an inert nuclear core, above which exists the valence (model) space, consisting of single-particle levels which differ based on the target nucleus in consideration. For each valence space, a range of shell model ``interactions" exist, {each determined by fitting the free parameters of the nuclear Hamiltonian to selected sets} 
of data (typically excitation and binding energies) from a range of nuclei. 
\\
\\
The $^{40}$Ar shell model OBDMEs were obtained from calculations performed in the  $sdpf$ valence space, consisting of single particle levels $1d_{5/2}$, $2s_{1/2}$, $1d_{3/2}$, $1f_{7/2}$, $2p_{3/2}$, $1f_{5/2}$ and $2p_{1/2}$ above an inert \textsuperscript{16}O core. We employed the shell model interactions SDPF-NR \cite{Nummela:2001xh}, SDPF-U \cite{Nowacki:2007uh}, EPQQM \cite{Kaneko:2011ss} and SDPF-MU \cite{Utsuno:2012qf}. The valence space truncation employed here restricts the protons to the $sd$ shell, where they are free to occupy any of the levels, whilst the neutrons completely fill the $sd$ shell and remain unrestricted in the $pf$ shell. The aforementioned four shell model interactions produce different OBDMEs, and can be used to quantify uncertainties due to nuclear structure modelling. 
\\
\\
A Mathematica package \cite{Anand:2013yka} has been developed by Anand \textit{et al.} to compute the nuclear form factors in the shell model approach, given values for the OBDMEs. We use this package to compute the form factors from the shell model and mean-field approaches, the latter requiring some adaptation as discussed in the next section. 
\\
\\
Anand \textit{et al.}'s code uses a spherical harmonic oscillator single-particle basis for the nucleon wave functions. In this regime, the computed form factors are expressed in the form $e^{-2y}p(y)$, where $p(y)$ is a polynomial with $y=(qb/2)^2$, $$b= {1}/{\sqrt{m_N \omega}} \approx \sqrt{41.467/(45A^{-1/3} - 25A^{-2/3})}~\mbox{fm}$$  is the harmonic oscillator length parameter, and $\omega$ is the oscillator frequency. This thus provides a direct momentum-dependent expression for the form factors, which can be used in the differential scattering cross section. \\

\section{The Mean-Field Approach\label{sec:MF}}

\subsection{The HFB Equations}

\noindent We have used the program \textsc{hfbtho} to compute the nuclear ground state wavefunction of \textsuperscript{40}{Ar} with the Hartree-Fock-Bogoliubov (HFB) theory. The HFB theory applies a mean-field approach to nuclear structure. It uses a {Bogoliubov} 
transformation from the canonical basis of single particle states to a quasiparticle basis that allows for efficient computation of nuclear states including an approximated pairing correlation \cite{RingSchuck}. See  \cite{HFBTHO1,HFBTHO2,HFBTHO3,HFBTHO4} for a complete description on the computational methods of the code. Here, we provide a brief summary of the approach. We have used the HFB theory with a Skyrme energy density functional (EDF) \cite{skyrme1956} to compute the nuclear ground state, though we note that \textsc{hfbtho} also includes the option for a finite-range Gogny EDF \cite{decharge1980} as a possible area for future investigation. 
\\
\\
As the ground-state of an even-even nucleus like $^{40}$Ar is invariant under time-reversal symmetry, all time-odd densities vanish. In this case, the Skyrme EDF is a function of  the particle local density  $\rho_q$ ($q=\{p,n\}$ for protons and neutrons, respectively),  the kinetic energy density $\tau_q$,  the spin current density $\mathbf{J}_q$, and   the pairing local density $\tilde{\rho}_q$ (see Ref.~\cite{HFBTHO1} for their definitions). 
The Skyrme EDF is then defined as 
\begin{align}
    E[\rho_q, \tau_q,\mathbf{J}_q,\tilde{\rho}_q] &= \int d^3 \mathbf{r} \, [\mathcal{H}(\mathbf{r}) + \tilde{\mathcal{H}}(\mathbf{r})],
\end{align}
where $\mathcal{H}$ and $\tilde{\mathcal{H}}$ are the mean-field and pairing energy densities, respectively. 
Their expressions are given by
\begin{widetext}
\begin{align}
    \mathcal{H}(\mathbf{r}) &= \frac{\hbar^2}{2m} \tau + \frac{1}{2}t_0 \left[ \left(1 + \frac{1}{2}x_0\right)\rho^2 - \left(\frac 1 2 + x_0 \right)\sum\limits_1 \rho_q^2    \right] \nonumber
    \\
    &+ \frac{1}{2}t_1 \left[\left(1 + \frac 1 2 x_1 \right) \rho \left(\tau - \frac{3}{4} \Delta\rho \right)  - \left( \frac{1}{2} + x_1 \right) \sum\limits_q \rho_q \left(\tau_q - \frac{3}{4} \Delta \rho_q\right)\right] \nonumber
    \\
    &+ \frac{1}{2} t_2 \left[ \left(1 + \frac{1}{2} x_2\right)\rho\left(\tau + \frac{1}{4} \Delta \rho\right) - \left(\frac{1}{2} + x_2\right) \sum\limits_q \rho_q \left(\tau_q + \frac{1}{4} \Delta \rho_q\right) \right] \nonumber
    \\
    &+ \frac{1}{12} t_3 \rho^{\alpha} \left[ \left(1 + \frac{1}{2} x_3\right) \rho^2 - \left(x_3 + \frac{1}{2}\right) \sum\limits_q \rho_q^2 \right] \nonumber
    \\
    &- \frac{1}{8}\left(t_1 x_1 + t_2 x_2\right) \sum\limits_{ij}\mathbf{J}_{ij}^2 + \frac{1}{8} \left(t_1 - t_2\right) \sum\limits_{q,ij} \mathbf{J}_{q,ij}^2 - \frac{1}{2}W_0\sum\limits_{ijk} \epsilon{ijk}\left[ \rho \nabla_k \mathbf{J}_{ij} + \sum\limits_q \rho_q \nabla_k \mathbf{J}_{q,ij} \right],
    \\ \nonumber
    \\
    \tilde{\mathcal{H}}(\mathbf{r}) &= V_0\left[1 - V_1 \left( \frac{\rho}{\rho_0} \right)^{\gamma} \right] \sum\limits_q \tilde{\rho}_q^2.
\end{align}
\end{widetext}

\noindent The parameters $t_i, x_i, \alpha,$ and $W_0$ are determined by the choice of Skyrme parametrisation which{---as with the shell model interactions---}are computed from numerical fits to particular sets of nuclear data. For this work, we use the SLy4 parametrisation \cite{CHABANAT1998231}. The parameters $V_i$ and $\gamma$ are separately fixed to determine the pairing characteristics of the computation \cite{HFBTHO1}. $\gamma=1,V_1 = 0.5$ corresponds to a mixed surface-volume pairing character. We have selected the same pairing strength $V_0= -310$ MeV {fm$^3$} 
for both protons and neutrons---and a pairing cutoff energy of 60 MeV---based on the work of Chen \textit{et al.} \cite{PairingFit}, who have performed a fit of the pairing strength for several Skyrme parametrisations and pairing forces to the experimentally observed neutron pairing gap of \textsuperscript{120}Sn. The values of all force parameters used are shown in Table~\ref{SLy4}. 
\\

\begin{table}
    \centering
    \begin{tabular}{l|l||l|l}
parameter & value & parameter & value\\
    \hline\hline
       $t_0$ (MeV fm\textsuperscript{3}) & -2488.913 & $x_0$  & 0.834 \\
    \hline
       $t_1$ (MeV fm\textsuperscript{5}) & 486.818 & $x_{{1}} 
       $ & -0.344 \\
    \hline
       $t_2$ (MeV fm\textsuperscript{5}) & -546.395 & $x_{{2}}$ 
       & -1 \\
    \hline
       $t_3$ (MeV fm\textsuperscript{3 + $3\sigma$}) & 13777 & $x_{{3}}$ 
       & 1.354\\
    \hline
       $W_0$ (MeV fm\textsuperscript{5}) & 123 & $\alpha$ & 1/6 \\
    \hline
       $V_0$ (MeV {fm$^3$}) 
       &  -310 & $\gamma$ & 1 \\
    \hline
       $\rho_0$ (fm$^{-3}$) & 0.160 & $V_1$ & 0.5 
    
    \end{tabular}
    \caption{SLy4 parameterisation \cite{CHABANAT1998231} of the Skyrme EDF completed with user-selected pairing parameters. 
    }
    \label{SLy4}
\end{table}

\noindent The HFB theory employs matrices $U$ and $V$ to perform a Bogoliubov transformation between the canonical basis $c_n$ of nuclear states and the quasiparticle basis $\alpha_k$:
\begin{align}
    \begin{pmatrix}
        \alpha
        \\
        \alpha^{\dagger}
    \end{pmatrix}
    &= \begin{pmatrix}
        U^{\dagger} & V^{\dagger}
        \\
        V^T & U^T
    \end{pmatrix}
    \begin{pmatrix}
        c
        \\
        c^{\dagger}
    \end{pmatrix}.
\end{align}
\noindent The HFB equations are then derived within the quasiparticle basis from variation of the EDF with respect to $\rho$ and $\tilde{\rho}$:
\begin{align}
    &\sum\limits_{\sigma'} \begin{pmatrix}
        h(\mathbf{r},\sigma,\sigma') & \tilde{h}(\mathbf{r},\sigma,\sigma') 
        \\
        \tilde{h}(\mathbf{r},\sigma,\sigma')  & - h(\mathbf{r},\sigma,\sigma') 
    \end{pmatrix}
    \begin{pmatrix}
        U(E,\mathbf{r}\sigma')
        \\
        V(E,\mathbf{r}\sigma')
    \end{pmatrix} \nonumber
    \\
    &= \begin{pmatrix}
        E + \lambda & 0 
        \\
        0 & E - \lambda
    \end{pmatrix}
    \begin{pmatrix}
        U(E,\mathbf{r}\sigma)
        \\
        V(E,\mathbf{r}\sigma)
    \end{pmatrix},
\end{align}
\noindent where $\sigma,\sigma'$ are nucleon spins, $h,\tilde{h}$ are local fields determined by the EDF (the definitions are omitted here for brevity, and can be found in \cite{HFBTHO1}) and $\lambda$ are the standard Lagrange multipliers imposed to fix the particle number. These derived Skyrme-HFB equations are then solved iteratively to determine the nuclear energy eigenvalues $E$ and the corresponding eigenstate wavefunctions. HFBTHO does so by expanding the quasiparticle wavefunctions in the harmonic oscillator basis, using either the standard basis (HO), or a transformed basis (THO) achieved by applying a local scaling transformation to the HO basis. In this work, we use the standard HO basis, which is sufficient to describe the spherical \textsuperscript{40}Ar. After a ground-state solution is found, a transformation back to the canonical basis is performed to extract the relevant physical quantities.
\\
\\
{The HO basis consists of the eigenfunctions for a (possible axially deformed) potential \cite{HFBTHO1}. 
However, when the HO potential is chosen to be spherically symmetric, as in our case, 
the basis functions become the usual spherical harmonic oscillator functions. To maintain consistency with the oscillator length employed by Anand \textit{et al.}'s code for computing the form factors, we have set $b_0 = \sqrt{41.467/(45A^{-1/3} - 25A^{-2/3})}~\mbox{fm}$. 
\\
\\
The results presented in this paper were obtained from a calculation including the first 6 shells of the harmonic oscillator in the computation, giving a total of 84 basis functions. The computation converged to a nuclear binding energy of $343.134$ MeV, providing good agreement with the NUBASE2020 recorded value of $343.8$ MeV \cite{Kondev_2021}. To study how the selected basis affects convergence, we have also performed computations iwith smaller number harmonic oscillator shells. For instance, a calculation with 4 shells  converged to a slightly lower binding energy of $341.027$ MeV respectively. For the purposes of our work, however, the question of interest is how the truncation of the basis alters the form factor used to compute physical cross-sections. This is most easily assessed through the integrated form factors presented in Sec.~\ref{sec: IFF results}, and we comment on this effect there.} 

\subsection{The Nuclear Density Matrix}

\noindent {The} key quantity 
to determine the nuclear form factors is the nuclear density matrix. For compatibility with the existing code of Anand \textit{et al.} used to compute the {form} 
factors in the shell-model, this requires the density matrix to be found in the spherical harmonic oscillator basis. HFBTHO provides as output the energies of the nucleons in the canonical basis as well as the overlap of their wavefunctions with the nucleon states identified by the Nilsson quantum numbers. For a spherical nucleus, the Nilsson states can be exactly identified with states in the spherical harmonic oscillator basis \cite{firestone_shirley_baglin_chu_zipkin_1997}. As such, the proof-of-principle computation presented in this paper for spherical \textsuperscript{40}Ar requires only the computation of the density matrix in the basis of Nilsson quantum numbers. 
\\
\\
Defining $U_{\alpha i} = \braket{\phi_i\lvert\varphi_{\alpha}}$ as the overlaps between canonical states $\alpha$ and Nilsson states $i$, the required density matrix is then:
\begin{align}
    \rho_{ij} &= \sum\limits_{\alpha} n_{\alpha} U^*_{\alpha j} U_{\alpha i},
\label{eq:roij}
\end{align}
where $n_{\alpha}$ is the (possibly fractional) occupation number of the canonical state $\alpha$.  
\\
\\
We now draw attention to the inclusion of a pairing interaction in our computations. The use of a pairing force breaks particle number symmetry. This can lead to inaccuracies in the computation of the form factors: since the nuclear wavefunction is now a superposition of different states with good particle number, states differing from the physical $Z$ and $N$ values also contribute to the density matrix. The solution to this is to perform a projection of the nuclear ground state onto the state with the correct particle number, and compute the form factors for this projected state. 

\subsection{Particle Number Projection}

\noindent Determining the wavefunction for the physically relevant state requires projection onto the corresponding particle numbers: $Z=18$ and $N=22$ in  this work. The particle number projection operator is defined as:
\begin{align}
    \hat{P}^{N_0} &= \frac{1}{2\pi} \int\limits_0^{2\pi} e^{i\varphi(\hat{N} - N_0)} d\varphi, 
\end{align}
where $\hat{N} = \sum\limits_n \hat{c}_n^{\dagger} \hat{c}_n $ is the nucleon number operator (applying to either protons or neutrons separately) summed over all nucleon states $n$, and $N_0$ is the number of nucleons to be fixed by the projection operator. To discretise this expression, HFBTHO employs the Fomenko expansion \cite{Fomenko}:
\begin{align}
    \hat{P}^{N_0} &\simeq \frac{1}{N_{\varphi}} \sum\limits_{\ell=1}^{N_{\varphi}} e^{i\varphi_{\ell}(\hat{N} - N_0)},
\end{align}
where $N_{\varphi}$ is the number of gauge angle points in the integration mesh, and the gauge points are given by $\varphi_{\ell} = \frac{\pi}{N_{\varphi}}\ell$. Note that HFBTHO's enforcement of parity symmetry means integration only needs to be carried over the interval $[0,\pi]$. Fixing an odd number of gauge angle points is useful to avoid a pole that can sometimes appear at $\varphi=\pi/2$ in the integration path \cite{PNPBog}, and for our purposes 15 points proved sufficient. 
\\
\\
Rather than needing the projected wavefunctions themselves, we require only the one-body density matrices for the projected state. These can be expressed in terms of the unprojected density matrices \cite{SHEIKH200071}. To illustrate this, we observe the action of the projection operator on the nuclear wavefunction:
\begin{align}
    \hat{P}^{N} \ket{\Psi} & = \ket{\Psi_N} ,
\end{align}
and we define the `transition states':
\begin{align}
    \ket{\Psi(\varphi_{{\ell}})} &:= e^{i\varphi_{\ell}\hat{N}}\ket{\Psi}. 
\end{align}
The projected state can then be expressed as a superposition of transition states:
\begin{align}
    \ket{\Psi_N} &\simeq \frac{1}{N_{\varphi}}\sum\limits_{\ell=1}^{N_{\varphi}} e^{-i\varphi_{{\ell}} N_0} \ket{\Psi(\varphi_{\ell})}.
\end{align}

\noindent The normalised projected density matrix elements will be:
\begin{align}
    \rho_{ij}^{N_0} &= \frac{\bra{\Psi}\hat{c}_j^{\dagger} \hat{c}_i \hat{P}^{N_0} \ket{\Psi}}{\bra{\Psi}\hat{P}^{N_0} \ket{\Psi}}, 
\end{align}
which can be re-written as:
\begin{align}
    \rho_{ij}^{N_0} &\simeq \frac{\sum\limits_{\ell=1}^{N_{\varphi}} e^{-i\varphi_{{\ell}} N_0}\rho_{ij}(\varphi_{{\ell}}) \braket{\Psi\lvert\Psi(\varphi_{{\ell}})}}{\sum\limits_{\ell=1}^{N_{\varphi}} e^{-i\varphi_{{\ell}} N_0} \braket{\Psi\lvert\Psi(\varphi_{{\ell}})}}, 
\end{align}
where $\rho(\varphi)$ are likewise the transition densities $\frac{\bra{\Psi}\hat{c}_j^{\dagger} \hat{c}_i \ket{\Psi(\varphi)}}{\braket{\Psi\lvert\Psi(\varphi)}}$, found from \cite{SHEIKH200071}:
\begin{align}
    \rho(\varphi) &= e^{2i\varphi}\left(1 + \rho\left(e^{2i\varphi} -1\right) \right)^{-1} {\rho}, 
\end{align}
where $\rho$ in the right-hand side is defined in Eq.~(\ref{eq:roij}).
\\
\\
The overlaps $\braket{\Psi\lvert\Psi(\varphi)}$ are computed by HFBTHO during the projection, and we have input these into the above expression to determine the density matrices for the particle-number-projected state. In our results, we present the form factors computed both before and after projection.

\subsection{The Nuclear Form Factors}

\noindent The elements of the nuclear density matrix can be written as:
\begin{align}\label{eq: MF output}
    \langle J_i M_{J_f} ; T M_T | a^\dagger_{\alpha} a_{\beta} | J_i M_{J_i} ; T M_T \rangle,
\end{align}

\noindent where the single-particle states are denoted by the quantum numbers $\beta= \{n_\beta, l_\beta, j_\beta, t_\beta=1/2, m_{j_\beta}, m_{t_\beta}\}$. Note that the matrix elements produced are for $m_{t_\alpha} = m_{t_\beta}$. Computing the form factors using the Mathematica package of Anand \textit{et al.} \cite{Anand:2013yka} requires us to input the reduced density matrix in the spherical harmonic oscillator basis as defined in Eq.~(\ref{eq:obdme}). For our test case with an even-even nucleus, we have $J_i=M_{J_i}=M_{J_f}=0$. This allows {a simplified expression for conversion to the appropriate form}: 

\begin{widetext}
\begin{equation}\label{eq:MFOBDME}
\begin{split}
    \Psi^{0;\tau}_{|\alpha|, |\beta|}  & \equiv  \frac{\langle 0; T \ \vdots \vdots \left[a^\dagger_{|\alpha|} \otimes \tilde{a}_{|\beta|} \right]_{0;\tau} \vdots \vdots \ 0 ; T \rangle}{\sqrt{(2\tau+1)}}  \\
    & = \sum_{\substack{m_{t_\beta}, m_{j_\beta}, m_{j_\alpha}}} (-1)^{T-M_T+j_\beta -m_{j_\beta}+1/2-m_{t_\beta}} 
    \frac{\langle 0 \ 0 ; T M_T \ | a^\dagger_{\alpha}  a_{\beta}  | \ 0 \ 0 ; T M_T \rangle}{\sqrt{(2\tau+1)}} \\
   & \frac{\langle j_\alpha \ m_{j_\alpha} ; j_\beta \ - m_{j_\beta} | 0 \ 0 \rangle \langle 1/2 \ m_{t_\beta} ; 1/2 \ - m_{t_\beta} | \tau \ 0 \rangle}{\begin{pmatrix}
         \vspace{2mm} T  &  \tau  &  T \\
         \vspace{2mm} -M_T &  0 &  M_T  \hspace{2mm}
         \end{pmatrix}}.
\end{split}
\end{equation}
\end{widetext}
\noindent The above gives a direct expression for the reduced density matrix in terms of the computed mean-field nuclear density matrix. The computation of the form factors is then a simple matter, utilising the expressions shown in the preceding section. 
{In particular, these OBDMEs are employed in Eq.~(\ref{reduced nuclear matrix element}) to obtain the form factors $F_{X,Y}^{\tau,\tau'}$, which can be converted to proton-neutron formalism through}
\vspace{-0.3cm}
\begin{equation}\label{eq: form factor isospin pn conversion}
\begin{split}
        F_{X,Y}^{(p,p)} = \frac{1}{4} \left(F_{X,Y}^{0,0} +F_{X,Y}^{0,1}+F_{X,Y}^{1,0}+F_{X,Y}^{1,1} \right),
        \\
         F_{X,Y}^{(p,n)} = \frac{1}{4} \left(F_{X,Y}^{0,0} -F_{X,Y}^{0,1}+F_{X,Y}^{1,0}-F_{X,Y}^{1,1} \right),
         \\
          F_{X,Y}^{(n,p)} = \frac{1}{4} \left(F_{X,Y}^{0,0} +F_{X,Y}^{0,1}-F_{X,Y}^{1,0}-F_{X,Y}^{1,1} \right),
          \\
        F_{X,Y}^{(n,n)} = \frac{1}{4} \left(F_{X,Y}^{0,0} -F_{X,Y}^{0,1}-F_{X,Y}^{1,0}+F_{X,Y}^{1,1} \right).
\end{split}
\end{equation}

\section{Results and Discussion\label{sec:results}}

\subsection{Single-Particle Occupation Numbers}\label{sec: occupation results}

\noindent As a measure of whether the density matrices computed from the HFBTHO output and our conversion in Eq.~(\ref{eq:MFOBDME}) provide reasonable OBDMEs, we employ these values to calculate the proton and neutron occupation numbers of the single-particle levels. 
\\
\\
Writing the occupation number $\mathcal{N}$ for each single-particle level defined by quantum numbers $\beta$ in terms of the OBDMEs $\Psi^{J;\tau}_{|\beta|, |\beta|}$, we have

\begin{widetext}
\begin{equation}
    \begin{split}
         \mathcal{N}(|\beta|; m_{t_\beta}) \equiv  & \ \sum_{m_{j_\beta}} \sum_{M_{J_f}, M_{J_i}} \frac{\langle J_i M_{J_f} ; T M_T | a_\beta^\dagger \ a_\beta | J_i M_{J_i} ; T M_T \rangle}{2J_i+1}  \\
         = & \sum_{m_{j_\beta}} \sum_{\substack{M_{J_f}, M_{J_i}
         \\ \tau, J, M}} (-1)^{J_i-M_{J_f}+T-M_T+j_\beta -m_{j_\beta} +1/2 -m_{t_\beta}} \frac{\sqrt{(2J+1) (2\tau+1)}}{2J_i+1} 
         \begin{pmatrix}
         \vspace{2mm} J_i  &  J  &  J_i \\
         \vspace{2mm} -M_{J_f} &  M &  M_{J_i}  \hspace{2mm}
         \end{pmatrix} \\
         & \times  \begin{pmatrix}
         \vspace{2mm} T  &  \tau  &  T \\
         \vspace{2mm} -M_T &  0 &  M_T  \hspace{2mm}
         \end{pmatrix}
         \langle j_\beta \ m_{j_\beta} ; j_\beta \ - m_{j_\beta} | J \ M \rangle  \langle 1/2 \ m_{t_\beta} ; 1/2 \ - m_{t_\beta} | \tau \ 0 \rangle \ \Psi^{J;\tau}_{|\beta|, |\beta|} .
    \end{split}
\end{equation}
\end{widetext}

\noindent Here, we have summed over all OBDMEs associated with the $\beta$ state, i.e., those with different $J$, $M$, and $\tau$. We take $|\beta|=\{n_\beta, l_\beta, j_\beta \}$. This gives a separate occupation number for each $j_\beta$ level with $m_{t_\beta} = -1/2$ (neutrons) and $m_{t_\beta}=1/2$ (protons). The OBDMEs employed for this calculation can be found in Appendix~\ref{appndx:OBDMEs}. \\

\begin{table*}[htpb]
\captionsetup{justification=Justified}
\caption{Proton ($n^p_{nlj}$) and neutron ($n^n_{nlj}$) occupation numbers for all single-particle levels up to and including $3d_{3/2}$, where the level labels are designated by the radial quantum number $n$ (beginning at 1), the nucleon orbital angular momentum $l$, and total angular momentum $j$. This is presented for the mean-field (MF) and mean-field projected (MF-P) calculations, alongside four shell model results (SDPF-NR, SDPF-U, SDPF-MU, EPQQM). The horizontal line separates the core states from the valence space in the shell-model calculations. 
\label{tab:OccupationNumbers}}
\centering
\resizebox{\textwidth}{!}{\begin{tabular}{lccc|ccc|ccc|ccc|ccc|ccc|cc}
\hline \hline 
 &  &  & & \multicolumn{2}{c}{SDPF-NR} & & \multicolumn{2}{c}{SDPF-U} & & \multicolumn{2}{c}{EPQQM} & & \multicolumn{2}{c}{SDPF-MU} & & \multicolumn{2}{c}{MF} & & \multicolumn{2}{c}{MF-P}  \\
\cline{5-6}
\cline{8-9}
\cline{11-12}
\cline{14-15}
\cline{17-18}
\cline{20-21}
$n$  & $l$ & $j$ & {\footnotesize Max Occ} & $n^p_{nlj}$ & $n^n_{nlj}$ & & $n^p_{nlj}$ & $n^n_{nlj}$ & & $n^p_{nlj}$ & $n^n_{nlj}$ & & $n^p_{nlj}$ & $n^n_{nlj}$ & & $n^p_{nlj}$ & $n^n_{nlj}$ & & $n^p_{nlj}$ &$n^n_{nlj}$ \\
\hline \hline
 1 & 0 & $1/2$  & 2  & 2    & 2    &  & 2    & 2    &  & 2    & 2    &  & 2    & 2    &  & 1.996303 & 1.997122 & & 1.996496 & 1.997437 \\
1 & 1 & $3/2$  & 4  & 4    & 4    &  & 4    & 4    &  & 4    & 4    &  & 4    & 4    &  & 3.993052 & 3.992457 & & 3.99458  & 3.99374  \\
1 & 1 & $1/2$  & 2  & 2    & 2    &  & 2    & 2    &  & 2    & 2    &  & 2    & 2    &  & 1.996871 & 1.996146 & &  1.997792 & 1.996964 \\
\hline
1 & 2 & $5/2$  & 6  & 5.92 & 6    &  & 5.91 & 6    &  & 5.92 & 6    &  & 5.92 & 6    &  & 5.470287 & 5.9531   & & 5.493431 & 5.970954 \\
2 & 0 & $1/2$  & 2  & 1.83 & 2    &  & 1.82 & 2    &  & 1.78 & 2    &  & 1.85 & 2    &  & 1.412168 & 1.972849 & & 1.407274 & 1.984578 \\
1 & 2 & $3/2$  & 4  & 2.25 & 4    &  & 2.27 & 4    &  & 2.23 & 4    &  & 2.23 & 4    &  & 3.007262 & 3.946869 & & 3.02954  & 3.968618 \\
1 & 3 & $7/2$  & 8  & 0    & 1.79 &  & 0    & 1.83 &  & 0    & 1.91 &  & 0    & 1.47 &  & 0.042784 & 1.334580 & & 0.022265 & 1.315569 \\
2 & 1 & $3/2$  & 4  & 0    & 0.11 &  & 0    & 0.08 &  & 0    & 0.06 &  & 0    & 0.4  &  & 0.012501 & 0.325326 & & 0.006499 & 0.318552 \\
1 & 3 & $5/2$  &  6  & 0    & 0.07 &  & 0    & 0.07 &  & 0    & 0.02 &  & 0    & 0.06 &  & 0.01514  & 0.348847 & & 0.007865 & 0.336904 \\
2 & 1 & $1/2$  & 2  & 0    & 0.02 &  & 0    & 0.02 &  & 0    & 0.01 &  & 0    & 0.07 &  & 0.003123 & 0.055648 & & 0.001626 & 0.052374 \\
1 & 4 & $9/2$  & 10 & -    & -    &  & -    & -    &  & -    & -    &  & -    & -    &  & 0.006957 & 0.01881  & & 0.004880 & 0.014919 \\
1 & 4 & $7/2$  & 8  & -    & -    &  & -    & -    &  & -    & -    &  & -    & -    &  & 0.002955 & 0.011333 & & 0.001782 & 0.009420 \\
2 & 2 & $5/2$  & 6  & -    & -    &  & -    & -    &  & -    & -    &  & -    & -    &  & 0.004916 & 0.007131 & & 0.004101 & 0.005595 \\
2 & 2 & $3/2$  & 4  & -    & -    &  & -    & -    &  & -    & -    &  & -    & -    &  & 0.002364 & 0.006897 & & 0.001910 & 0.006364 \\
3 & 0 & $1/2$  & 2  & -    & -    &  & -    & -    &  & -    & -    &  & -    & -    &  & 0.005933 & 0.002524 & & 0.005852 & 0.002097 \\
1 & 5 & $11/2$ & 12 & -    & -    &  & -    & -    &  & -    & -    &  & -    & -    &  & 0.002309 & 0.005010 & & 0.001600 & 0.004049 \\
1 & 5 & $9/2$  & 10 & -    & -    &  & -    & -    &  & -    & -    &  & -    & -    &  & 0.001011 & 0.004112 & & 0.000568 & 0.003472 \\
2 & 3 & $5/2$  & 6  & -    & -    &  & -    & -    &  & -    & -    &  & -    & -    &  & 0.000845 & 0.002470 & & 0.000558 & 0.002181 \\
2 & 3 & $7/2$  & 8  & -    & -    &  & -    & -    &  & -    & -    &  & -    & -    &  & 0.002087 & 0.002322 & & 0.001561 & 0.001819 \\
3 & 1 & $1/2$  & 2  & -    & -    &  & -    & -    &  & -    & -    &  & -    & -    &  & 0.000930 & 0.000703 & & 0.000841 & 0.000620 \\
3 & 1 & $3/2$  & 4  & -    & -    &  & -    & -    &  & -    & -    &  & -    & -    &  & 0.002212 & 0.001221 & & 0.001989 & 0.000990 \\
1 & 6 & $13/2$ & 14 & -    & -    &  & -    & -    &  & -    & -    &  & -    & -    &  & 0.003186 & 0.003087 & &0.002877 & 0.002638 \\
2 & 4 & $9/2$  & 10 & -    & -    &  & -    & -    &  & -    & -    &  & -    & -    &  & 0.000709 & 0.001102 & & 0.000540 & 0.000883 \\
3 & 2 & $5/2$  & 6  & -    & -    &  & -    & -    &  & -    & -    &  & -    & -    &  & 0.000931 & 0.001028 & & 0.000801 & 0.000869 \\
1 & 6 & $11/2$ & 12 & -    & -    &  & -    & -    &  & -    & -    &  & -    & -    &  & 0.003513 & 0.003049 & & 0.003209 & 0.002643 \\
2 & 4 & $7/2$  & 8  & -    & -    &  & -    & -    &  & -    & -    &  & -    & -    &  & 0.005132 & 0.003215 & & 0.004954 & 0.002982 \\
4 & 0 & $1/2$  & 2  & -    & -    &  & -    & -    &  & -    & -    &  & -    & -    &  & 0.000629 & 0.000520 & & 0.000576 & 0.000462 \\
3 & 2 & $3/2$  & 4  & -    & -    &  & -    & -    &  & -    & -    &  & -    & -    &  & 0.004096 & 0.002269 & & 0.004071 & 0.002185\\
  \hline \hline
\end{tabular}}
\end{table*}

\noindent The resulting occupation numbers for the test-case mean-field model, as well as four different shell models, are provided in Table~\ref{tab:OccupationNumbers}. The top subtable presents these values for the core shell model levels, whilst the bottom subtable shows the valence levels. We note that this classification of ``core" and ``valence" levels is a shell model approach to nuclear structure, where the nucleons are restricted to stay within and below the valence shells. The mean-field calculation does not impose such a restriction, and includes OBDMEs up to and including the level $3d_{3/2}$ (noting that we did not consider higher levels in our calculation, as these are largely unoccupied). The total number of protons and neutrons present in the nucleus based on the unprojected MF occupation values  are $N_p= 18.0002$ and $N_n= 21.9997$, respectively, with the mass number coming to $39.9999$. For the projected states, the numbers are instead $N_p=18.0000$, $N_n=21.9999$, and $A = 39.9999$.  This is consistent within rounding with the expected nucleon numbers for $^{40}$Ar of $Z=18$ and $N=22$.
We note that the projected and unprojected MF density matrices are quite similar to each other, with the largest difference in matrix elements being $\sim 10^{-2}$.

\subsection{Form Factors}\label{sec: form factor results}

\noindent Using the mean field and projected mean field (MF-P) OBDMEs (provided in Appendix~\ref{appndx:OBDMEs}), the form factors are calculated using Eqs. \ref{Eq:FF} and \ref{reduced nuclear matrix element}, with the expressions included in \autoref{appendix:FF}.  
These can be compared to the shell model counterparts for the interactions SDPF-NR, SDPF-U, SDPF-MU and EPQQM in Refs.~\cite{AbdelKhaleq:2023ipt,AbdelKhaleq:2024hir}.

\begin{figure*}[htbp]
\centering 
\includegraphics[width=0.4\linewidth]{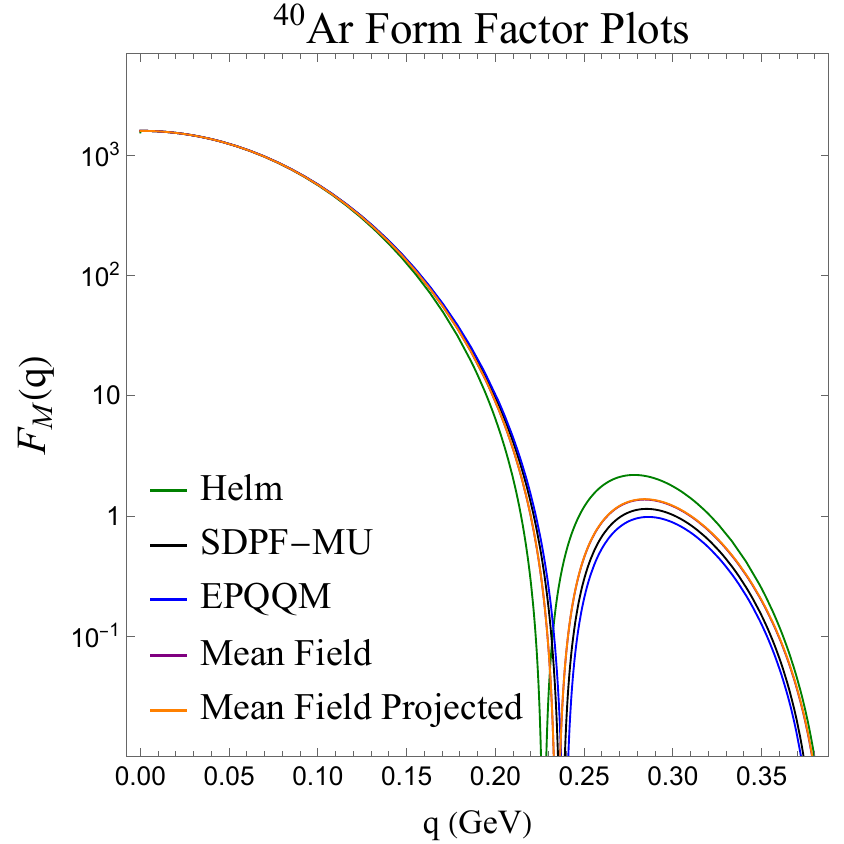}
\includegraphics[width=0.4\linewidth]{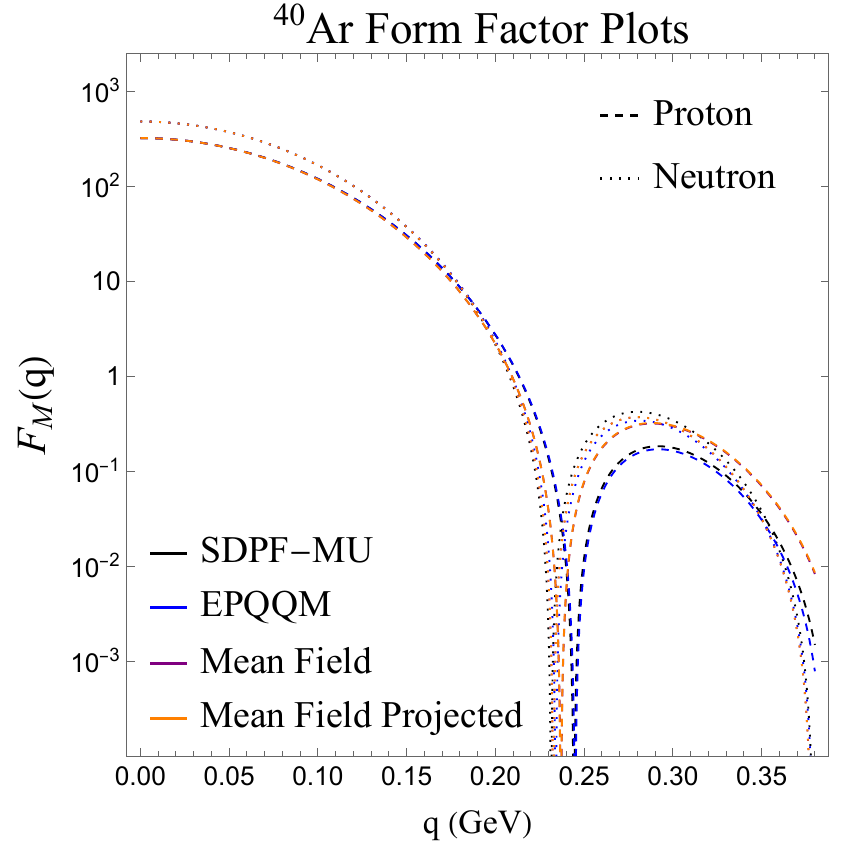}
\captionsetup{justification=Justified}
\caption{$^{40}$Ar form factors $M$  comparing shell model and mean-field results. The left figure shows full form factors presented against the Helm response, whilst the right figure shows the proton and neutron form factor contributions for select calculations. 
The unprojected and projected mean-field lines overlap each other. 
\label{fig:FFPlotM} }
\end{figure*}

\begin{figure*}[htbp]
\centering 
\includegraphics[width=0.4\linewidth]{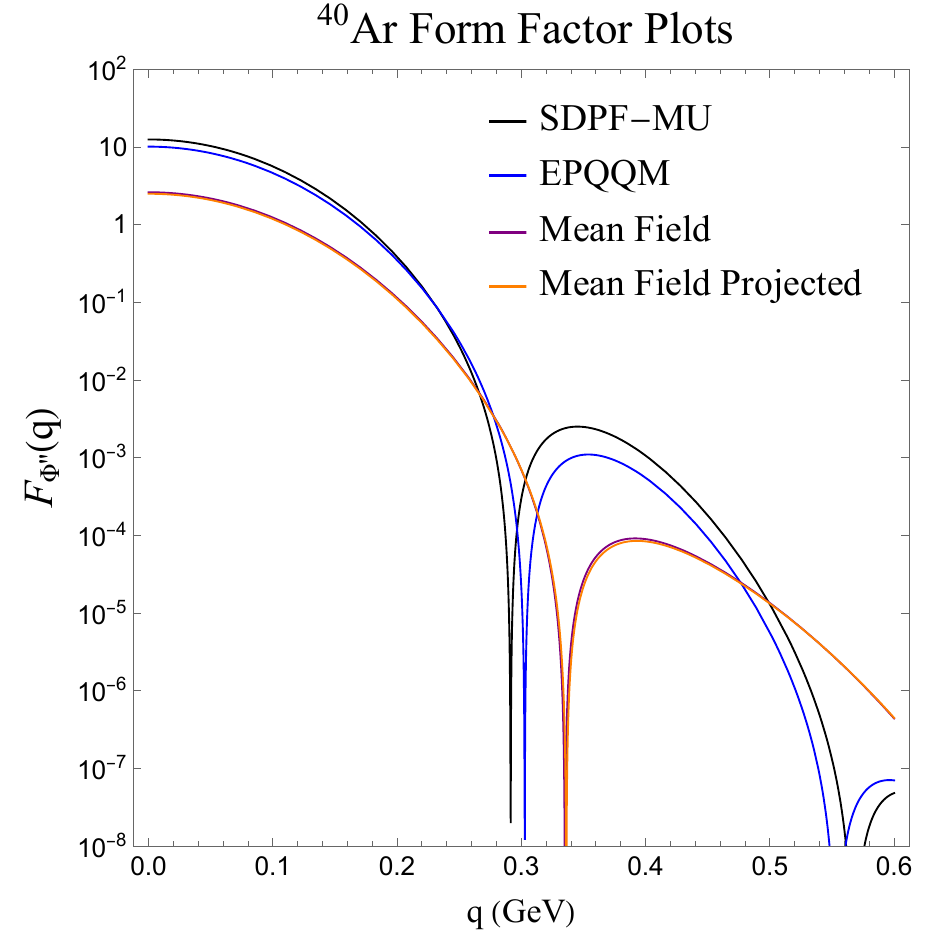}
\includegraphics[width=0.4\linewidth]{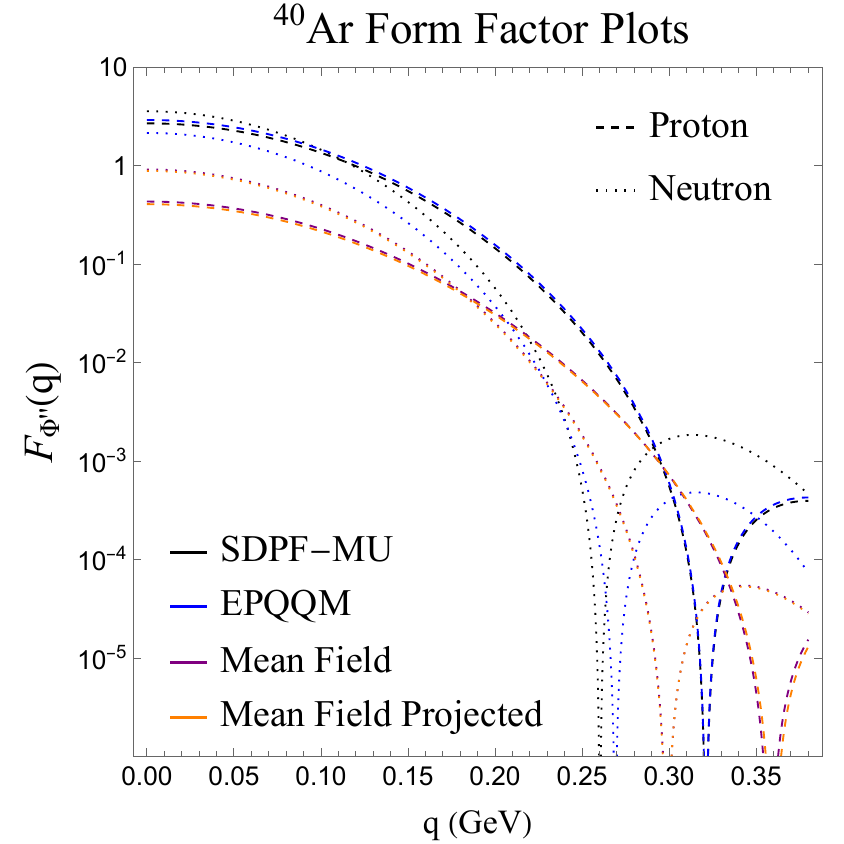}
\captionsetup{justification=Justified}
\caption{$^{40}$Ar form factors $\Phi''$  comparing shell model and mean-field results. The left figure shows full form factors presented against the Helm response, whilst the right figure shows the proton and neutron form factor contributions for select calculations. The unprojected and projected mean-field lines track each other closely, with the difference between the two most clearly seen in the right-hand plot. 
\label{fig:FFPlotPhi}}
\end{figure*}

\noindent To better understand the behaviour of these form factors over a range of momentum transfer $q$ values, they are plotted against the shell model counterparts in Fig.~\ref{fig:FFPlotM} for the $M$ channel and Fig.~\ref{fig:FFPlotPhi} for the $\Phi''$ channel. 
The left figure in each panel presents the full form factor $\sum_{N,N'=p,n} F^{(N,N')}_{X,X} (q^2)$ for each nuclear model. Additionally, the Helm form factor \cite{PhysRev.104.1466} is provided against the $M$ plots for comparison,
{where we have employed $F_{\rm Helm}(q) = 3 \frac{j_1(qR)}{qR} \ e^{-(qs)^2/2}$, with $j_1 (x)$ being the spherical Bessel function of the first kind and the nuclear radius $R$ has the form $R^2 =  c^2 + \frac{7}{3} \pi^2 a^2 -5s^2$ ($a \simeq 0.52$ fm, $s\simeq 0.9$ fm, $c \simeq (1.23 A^{1/3} -0.6)$ fm). The Helm form factor was normalised using $A^2 F_{\rm Helm}^2 (q)$ for consistency with our form factor plots.}
The right figure in each panel provides a decomposition into the proton $F^{(p,p)}_{X,X}$ (dashed) and neutron $F^{(n,n)}_{X,X}$ (dotted) components of the total form factor. These are presented for the mean-field and projected mean-field calculations against two shell model interactions, SDPF-MU and EPQQM, which display the largest nuclear differences between any two $^{40}$Ar shell model calculations performed here. \\

\noindent The theoretical SI form factors for $M$ are consistent with one another and with the Helm result up to $q \sim 160$ MeV. In particular, the mean-field and shell model calculations produce similar results for the entire $q$ range presented. This is expected for the SI channel where $M$ tends to the particle number when $q\rightarrow0$. Below $q \sim 180$ MeV the proton $M$ calculations agree perfectly, but beyond this $q$ the difference between the mean-field result and shell model results is larger than the disagreements between the different shell models, indicating an additional uncertainty stemming from the choice of nuclear structure model. 
\\
\\
The spin-orbit $\Phi''$ channel in Fig.~\ref{fig:FFPlotPhi} displays much larger differences between the mean-field and shell model results compared to the aforementioned $M$ plots. All four shell model calculations are strongly consistent with one another up to $q \sim 250$ MeV, and are still somewhat consistent beyond this point. The mean-field calculations on the other hand differs from the others even at low $q$ values, where at $q \sim 0$ the factor difference between shell model and mean-field is $\sim 4$. This is also reflected in the proton and neutron components, where the factor difference between the shell model and mean field $F^{(p,p)}$ ($F^{(n,n)}$) values at small $q$ is $\sim 7$ ($\sim 3$). \\
\\
\\
This difference in $\Phi''$ values likely comes from the nucleon single-particle level occupation numbers. The spin-orbit operator is largest when one spin-orbit partner is filled whilst the other is empty. In the shell model case, the protons are restricted to the $sd$ shell, and hence no proton contribution from the $pf$ shell is present, while the reverse is true for the neutrons. The high $\Phi''$ value in this case is likely due to the partial occupation of the proton $1d_{5/2,3/2}$ spin-orbit partners and partial occupation of the neutron $1f_{7/2,5/2}$ partners. 
\\
\\
The mean-field case is more complex, since nucleons are free to occupy any levels up to $3d_{3/2}$. Compared to the shell model case, the proton distribution over the $1d_{5/2,3/2}$ levels is more even, reducing the contribution to $\Phi''$ from these levels. The distribution of the nucleons across higher-energy single-particle levels, where the occupancy values are also very small in magnitude, likely leads to the smaller spin-orbit $\Phi''$ contribution in the mean-field case. 
\\
\\
Finally, the projected and unprojected MF form factors are in very close agreement, showing almost no visible difference in the plots. This is expected given the very similar density matrices, and indicates that at least for \textsuperscript{40}Ar the impact of the number-breaking pairing interaction does not significantly affect the form factors.

\subsection{Integrated Form Factors (IFFs)}\label{sec: IFF results}

\begin{figure*}[htbp]
\centering
\includegraphics[width=0.45\linewidth]{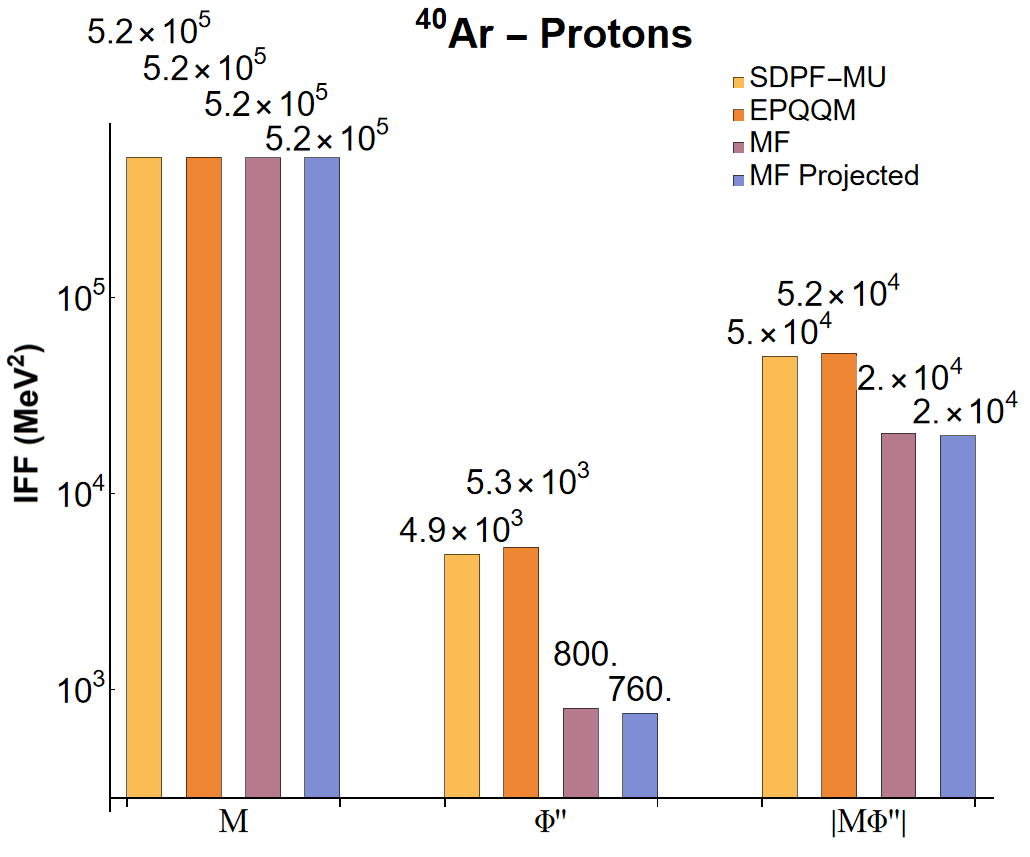}
\includegraphics[width=0.45\linewidth]{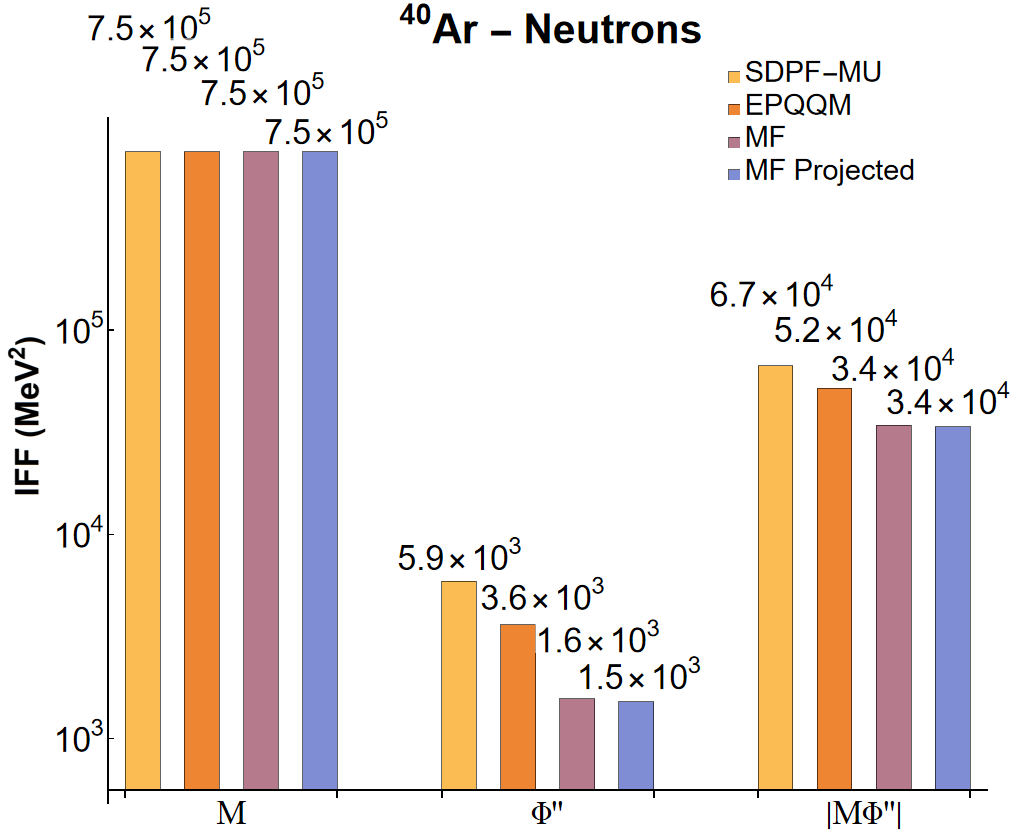}
\captionsetup{justification=Justified}
\caption{$^{40}$Ar proton and neutron IFF values for two shell model calculations (SDPF-MU and EPQQM) compared to mean-field models. The values are obtained for an integral with upper limit $q_{\rm max}=100$ MeV, and presented for the nuclear channels $M$, $\Phi''$, and their cross-term $|M\Phi''|$. 
\label{fig:IFF100}}
\end{figure*}

\noindent A standard approach in probing the nuclear response is to integrate the aforementioned form factors to obtain a momentum-integrated form factor (IFF) value for each nuclear channel $X$. These IFF values act as a gauge for the strength of each of the nuclear responses, which depend on the unique nuclear structure of the isotope in consideration. Following Refs.~\cite{Fitzpatrick:2012ix, AbdelKhaleq:2023ipt}, the proton and neutron IFF  are defined as
\begin{equation}\label{IFF expression}
  \int_{0}^{\text{100\text{ MeV}}} \frac{q \,dq}{2} F^{(N, N)}_{X(,Y)} (q^2),
\end{equation}
in units of $\text{MeV}^2$. The $^{40}$Ar proton and neutron IFF values are presented in Fig.~\ref{fig:IFF100}. The mean field result is compared to two shell model calculations, SDPF-MU and {EPQQM} 
\\

\noindent The SI $M$ response shows no difference between the shell model and mean-field calculations, consistent with the form factors plotted in Section~\ref{sec: form factor results}. The proton and neutron mean field $\Phi''$ IFFs are smaller than the shell model counterparts, also in line with the behaviour observed in Fig.~\ref{fig:FFPlotPhi}. The $\Phi''$ IFF factor difference of $\sim 7$ ($\sim 3$) for protons (neutrons) reflects the differences seen in the form factors at low $q$. This disagreement highlights the importance of choosing an appropriate nuclear model for a given nucleus employed in direct detection, in order to ensure an accurate calculation of the particular nuclear responses to an interaction.
\\
\\
\noindent The 100 MeV integral upper limit in Eq.~(\ref{IFF expression}) has been typical to use for all nuclei considered in such an analysis. However, a more realistic range of momentum transfer $q$ values is typically modelled after the range of interest for the particular DM direct detection experiment and associated nuclei being considered. Given that we do not consider a specific DM experiment in this work, we obtain this more realistic upper limit by considering a momentum transfer of $q \approx m_T v$ (for $m_\chi \sim m_T$) with $v=500$ km/s, giving a new upper limit of $q_{\rm max} \sim 63$ MeV. The updated IFF values using this new integral limit displayed factor differences between the nuclear models consistent with those seen above. This shows the upper limit $q_{\rm max}$ not to be of major significance to the nuclear uncertainty quantification for an $^{40}$Ar target  -- however, this analysis should be repeated for specific $q$ ranges corresponding to particular DM experiments. 
\\
\\
{Finally, we note how the choice of basis in the mean-field computation alters the IFFs. Comparing the 4- and 6-shell calculations lead to changes in IFF smaller than 6\%. the largest change occurs in the $\Phi''$ channel, while the change in the $M$ channel is negligible. 
The largest changes occurring in the $\Phi''$ channel is as expected: comparing the shell model and mean-field calculations shows this channel is the most sensitive to nuclear structure effects.} 

\section{Conclusions\label{sec:conclusions}}

\noindent In this work, we present a first investigation into the use of a mean-field model of nuclear structure in quantifying the nuclear response in dark matter direct detection experiments. This contrasts with the traditional approach of applying a nuclear shell model in this context. We compute the nuclear ground-state wavefunction for \textsuperscript{40}Ar in the Skyrme-HFB theory, and use this result to compute the nuclear density matrix and resulting DM interaction form factors. We find that there can be significant deviation in the nuclear response between the shell model and mean-field model. This highlights the need to carefully choose an appropriate model for a given nucleus in question.
\\
\\
In this paper we have restricted attention to a spherical and doubly-even nucleus. An additional benefit of employing a mean-field model, however, is in its greater accuracy with deformed and odd nuclei. Appropriately computing all of the form factors for such nuclei requires additional methods of angular momentum projection after computation of the nuclear ground state, which we will address in a future paper.

\section*{Acknowledgements}
\noindent This research was supported by the Australian Government through the Australian Research Council Centre of Excellence for Dark Matter Particle Physics (CDM, CE200100008). 

\section*{Appendices}

\appendix

\section{One-Body Density Matrix Elements (OBDMEs)}\label{appndx:OBDMEs}

\noindent The OBDMEs employed in the calculation of the form factors are displayed in Tables~\ref{tab:tau=03Shell}, \ref{tab:tau=13Shell} and \ref{tab:OBDM6Shell}. We note that for the shell model core levels, the OBDME is the same regardless of shell model interaction employed, as the core is assumed to be completely filled. Due to this, there are also no $\tau=1$ values in the shell model case. This is unlike the mean-field results, where $\tau=1$ OBDME values exist, although are small in magnitude compared to the $\tau=0$ values. The mean-field OBDMEs are shown against four shell model counterparts: SDPF-NR, SDPF-U, SDPF-MU and EPQQM. Unlike in the shell model case where the nucleons are constrained to the $sdpf$ valence space, the mean field calculation includes levels up to {$3d_{3/2}$} for both protons and neutrons, resulting in additional non-zero OBDMEs. {Additionally, the restrictions on the valence space for the shell model calculations prevent non-zero values for any off-diagonal matrix elements between states across different major shells (that is, when $N_{\text{out}} \neq N_{\text{in}})$. This restriction does not apply to the mean-field case, and so such off-diagonal matrix elements appear. Note that off-diagonal elements are symmetric under swaps $(N_{\rm{out}},j_{\alpha}) \leftrightarrow (N_{\rm{in}},j_{\beta})$, and so we omit redundant elements. We note also that since $J=0$, the matrix elements also vanish for $j_{\alpha} \neq j_{\beta}$, meaning off-diagonal matrix elements within a single shell do not emerge in any calculation.} 
While the OBDMEs between shell model and mean-field calculations are similar in the core levels, they tend to differ more significantly in the valence space.

\begin{widetext}

\begin{table}[H]
\resizebox{\textwidth}{!}{\begin{tabular}{|l|c|c|c|c|c|c|c|c|}
\hline
 $N_{\text{out}}$& $N_{\text{in}}$ & $j_{\alpha/\beta}$ & \hspace{1mm} $\Psi^{J;\tau}_{|\alpha|, |\beta|}$ (NR) & \hspace{1mm} $\Psi^{J;\tau}_{|\alpha|, |\beta|}$ (U) & \hspace{1mm} $\Psi^{J;\tau}_{|\alpha|, |\beta|}$(MU) & \hspace{1mm} $\Psi^{J;\tau}_{|\alpha|, |\beta|}$ (EP) & $\Psi^{J;\tau}_{|\alpha|, |\beta|}$ (MF) & $\Psi^{J;\tau}_{|\alpha|, |\beta|}$ (MF-P) \\
\hline
 \textbf{0}                & \textbf{0}               & \textbf{1}                  & 4.472136                                                               & 4.472136                                                              & 4.472136                                                                & 4.472136                                                               & 4.464785                                 & 4.465353                                   \\
 0                & 2               & 1                  & -                                                                      & -                                                                     & -                                                                       & -                                                                      & 0                                 & 0.000008                                   \\
 0                & 4               & 1                  & -                                                                      & -                                                                     & -                                                                       & -                                                                      & -0.060581                                & -0.060625                                  \\
 0                & 6               & 1                  & -                                                                      & -                                                                     & -                                                                       & -                                                                      & 0                                 & $3\times 10^{-7}$                          \\
  \textbf{1}                & \textbf{1}               & \textbf{1}                  & 4.472136                                                               & 4.472136                                                              & 4.472136                                                                & 4.472136                                                               & 4.464329                                 & 4.466272                                   \\
  1                & 3               & 1                  & -                                                                      & -                                                                     & -                                                                       & -                                                                      & 0.053211                                 & 0.053317                                   \\
 1                & 5               & 1                  & -                                                                      & -                                                                     & -                                                                       & -                                                                      & -0.064410                                & -0.064438                                  \\
\textbf{1}                & \textbf{1}               & \textbf{3}                  & 6.324555                                                               & 6.324555                                                              & 6.324555                                                                & 6.324555                                                               & 6.313099                                 & 6.315321                                   \\
 1                & 3               & 3                  & -                                                                      & -                                                                     & -                                                                       & -                                                                      & -0.000178                                & -0.000168                                  \\
 1                & 5               & 3                  & -                                                                      & -                                                                     & -                                                                       & -                                                                      & -0.067745                                & -0.067772                                  \\

 2                & 0               & 1                  & -                                                                      & -                                                                     & -                                                                       & -                                                                      & 0                                 & 0.000008                                   \\
 2                & 2               & 1                  & 4.278                                                                  & 4.272                                                                 & 4.226                                                                   & 4.306                                                                  & 3.784565                                 & 3.792205                                   \\
 2                & 4               & 1                  & -                                                                      & -                                                                     & -                                                                       & -                                                                      & -0.028573                                & -0.028644                                  \\
 2                & 6               & 1                  & -                                                                      & -                                                                     & -                                                                       & -                                                                      & -0.034208                                & -0.034172                                  \\
 2                & 2               & 3                  & 4.942                                                                  & 4.954                                                                 & 4.981                                                                   & 4.927                                                                  & 5.497723                                 & 5.532530                                   \\
 2                & 4               & 3                  & -                                                                      & -                                                                     & -                                                                       & -                                                                      & -0.089601                                & -0.089695                                  \\
 2                & 6               & 3                  & -                                                                      & -                                                                     & -                                                                       & -                                                                      & -0.107812                                & -0.108877                                  \\
 2                & 2               & 5                  & 7.696                                                                  & 7.690                                                                  & 7.694                                                                   & 7.692                                                                  & 7.373765                                 & 7.400229                                   \\
 2                & 4               & 5                  & -                                                                      & -                                                                     & -                                                                       & -                                                                      & 0.021294                                 & 0.021004                                   \\
 2                & 6               & 5                  & -                                                                      & -                                                                     & -                                                                       & -                                                                      & -0.038492                                & -0.038669                                  \\
 3                & 3               & 1                  & 0.02680                                                                 & 0.02327                                                               & 0.008617                                                                & 0.07809                                                                & 0.065708                                 & 0.060374                                   \\
 3                & 5               & 1                  & -                                                                      & -                                                                     & -                                                                       & -                                                                      & 0.002288                                 & 0.001928                                  
             \\
 3                & 3               & 3                  & 0.09054                                                                & 0.06587                                                               & 0.04396                                                                 & 0.3165                                                                 & 0.267076                                 & 0.256976                                   \\
 3                & 5               & 3                  & -                                                                      & -                                                                     & -                                                                       & -                                                                      & 0.005103                                 & 0.005299                                   \\
 3                & 3               & 5                  & 0.04715                                                                & 0.04554                                                               & 0.01527                                                                 & 0.04032                                                                & 0.234953                                 & 0.222548                                   \\
 3                & 5               & 5                  & -                                                                      & -                                                                     & -                                                                       & -                                                                      & 0.006952                                 & 0.006706                                   \\
 3                & 3               & 7                  & 0.9998                                                                 & 1.020                                                                  & 1.069                                                                   & 0.8203                                                                 & 0.769970                                 & 0.747872                                   \\
 3                & 5               & 7                  & -                                                                      & -                                                                     & -                                                                       & -                                                                      & 0.007809                                 & 0.006694                                   \\

\hline
\end{tabular}}
\captionsetup{justification=Justified}
\caption{{The OBDME values for $N_{\rm{out}} = 1,2,3$, for $J=0$ and $\tau=0$. The mean field (MF) and particle number projected mean-field (MF-P) results are compared against four different shell model calculations, SDPF-NR (NR), SDPF-U (U), SDPF-MU (MU) and EPQQM (EP). Here, $\beta$ ($\alpha$) represents the incoming (outgoing) state, with $N$ being the  harmonic oscillator shell number. Dashes indicate elements that vanish due to restrictions on the model space of shell model calculations. Bolded numbers indicate diagonal OBDMEs for core states in the shell model.} 
\label{tab:tau=03Shell}}
\end{table}

\begin{table}[H]
\resizebox{\textwidth}{!}{\begin{tabular}{|l|c|c|c|c|c|c|c|c|}
\hline
 $N_{\text{out}}$& $N_{\text{in}}$ & $j_{\alpha/\beta}$ & \hspace{1mm} $\Psi^{J;\tau}_{|\alpha|, |\beta|}$ (NR) & \hspace{1mm} $\Psi^{J;\tau}_{|\alpha|, |\beta|}$ (U) & \hspace{1mm} $\Psi^{J;\tau}_{|\alpha|, |\beta|}$(MU) & \hspace{1mm} $\Psi^{J;\tau}_{|\alpha|, |\beta|}$ (EP) & $\Psi^{J;\tau}_{|\alpha|, |\beta|}$ (MF) & $\Psi^{J;\tau}_{|\alpha|, |\beta|}$ (MF-P) \\
\hline
 \textbf{0}                & \textbf{0}               & \textbf{1}                  & -        & -        & -        & -        & 0.000647  & 0.000745          \\
 0                & 2               & 1                  & -        & -        & -        & -        & 0.000000  & -0.000004         \\
 0                & 4               & 1                  & -        & -        & -        & -        & 0.042837  & 0.042867          \\
 0                & 6               & 1                  & -        & -        & -        & -        & 0         & $4\times 10^{-7}$ \\
 \textbf{1}                & \textbf{1}               & \textbf{1}                  & -        & -        & -        & -        & -0.000573 & -0.000655         \\
 1                & 3               & 1                  & -        & -        & -        & -        & 0.044162  & 0.044248          \\
 1                & 5               & 1                  & -        & -        & -        & -        & 0.012849  & 0.012858          \\
 \textbf{1}                & \textbf{1}               & \textbf{3}                  & -        & -        & -        & -        & -0.000333 & -0.000469 \\
 1                & 3               & 3                  & -        & -        & -        & -        & -0.001492 & -0.001485         \\
 1                & 5               & 3                  & -        & -        & -        & -        & 0.039971  & 0.039988          \\
 2               & 0               & 1                  & -        & -        & -        & -        & 0         & -0.000004         \\
 2                & 2               & 1                  & 0.1372   & 0.1418   & 0.1742   & 0.1177   & 0.443257  & 0.456399          \\
 2                & 4               & 1                  & -        & -        & -        & -        & 0.007614  & 0.007386          \\
 2                & 6               & 1                  & -        & -        & -        & -        & 0.013980  & 0.014016          \\
 2                & 2               & 3                  & 0.9776   & 0.9690   & 0.9499   & 0.9881   & 0.525256  & 0.524960          \\
 2                & 4               & 3                  & -        & -        & -        & -        & -0.023165 & -0.023687         \\
 2                & 6               & 3                  & -        & -        & -        & -        & 0.030634  & 0.031379          \\
 2                & 2               & 5                  & 0.03550  & 0.03978  & 0.03668  & 0.03814  & 0.220373  & 0.217958          \\
 2                & 4               & 5                  & -        & -        & -        & -        & 0.022590  & 0.022999          \\
 2                & 6               & 5                  & -        & -        & -        & -        & -0.001451 & -0.001428         \\
 3                & 3               & 1                  & 0.01895  & 0.01646  & 0.006093 & 0.05522  & 0.041525  & 0.040120          \\
 3                & 5               & 1                  & -        & -        & -        & -        & 0.000736  & 0.000858 \\
 3                & 3               & 3                  & 0.06402  & 0.04658  & 0.03108  & 0.2238   & 0.174875  & 0.174443          \\
 3                & 5               & 3                  & -        & -        & -        & -        & 0.004585  & 0.004261          \\
 3                & 3               & 5                  & 0.03334  & 0.03220  & 0.01080  & 0.02851  & 0.152316  & 0.150185          \\
 3                & 5               & 5                  & -        & -        & -        & -        & 0.004877  & 0.004728          \\
 3                & 3               & 7                  & 0.7070   & 0.7215   & 0.7562   & 0.5800   & 0.510627  & 0.511223          \\
 3                & 5               & 7                  & -        & -        & -        & -        & 0.002315  & 0.003052          \\
\hline
\end{tabular}}
\captionsetup{justification=Justified}
\caption{{Same as Tab.~\ref{tab:tau=03Shell} for $\tau=1$. }
\label{tab:tau=13Shell}}
\end{table}

\begin{table}[H]
\centering
\begin{tabular}{|l|c|c|c|c|c|c|}
\hline
 $N_{\text{out}}$& $N_{\text{in}}$ & $j_{\alpha/\beta}$ & $\Psi^{J;\tau=0}_{|\alpha|, |\beta|}$ (MF) & $\Psi^{J;\tau=0}_{|\alpha|, |\beta|}$ (MF-P) & $\Psi^{J;\tau=1}_{|\alpha|, |\beta|}$ (MF) & $\Psi^{J;\tau=1}_{|\alpha|, |\beta|}$ (MF-P) \\
\hline
4 & 4 & 1  & 0.009455  & 0.008888  & -0.002694 & -0.002969 \\
4 & 6 & 1  & 0.000245  & 0.000152  & -0.000152 & -0.000244 \\
4 & 4 & 3  & 0.007322  & 0.006542  & 0.002534  & 0.002490  \\
4 & 6 & 3  & 0.001519  & 0.001393  & 0.000472  & 0.000439  \\
4 & 4 & 5  & 0.007776  & 0.006259  & 0.001011  & 0.000682  \\
4 & 6 & 5  & 0.000092  & -0.000028 & 0.000377  & 0.000254  \\
4 & 4 & 7  & 0.007987  & 0.006262  & 0.003312  & 0.003019  \\
4 & 6 & 7  & 0.000494  & 0.000258  & 0.000436  & 0.000273  \\
4 & 4 & 9  & 0.012884  & 0.009899  & 0.004191  & 0.003549  \\
4 & 6 & 9  & 0.000153  & 0.000031  & 0.000181  & 0.000163  \\
5 & 5 & 1  & 0.001826  & 0.001633  & -0.000180 & -0.000175 \\
5 & 5 & 3  & 0.002714  & 0.002355  & -0.000554 & -0.000559 \\
5 & 5 & 5  & 0.002140  & 0.001768  & 0.000742  & 0.000741  \\
5 & 5 & 7  & 0.002465  & 0.001889  & 0.000093  & 0.000102  \\
5 & 5 & 9  & 0.002561  & 0.002020  & 0.001096  & 0.001027  \\
5 & 5 & 11 & 0.003341  & 0.002578  & 0.000872  & 0.000790  \\
6 & 6 & 1  & 0.001285  & 0.001161  & -0.000086 & -0.000090 \\
6 & 6 & 3  & 0.005032  & 0.004945  & -0.001021 & -0.001054 \\
6 & 6 & 5  & 0.001264  & 0.001078  & 0.000045  & 0.000031  \\
6 & 6 & 7  & 0.004666  & 0.004436  & -0.000758 & -0.000779 \\
6 & 6 & 9  & 0.000905  & 0.000711  & 0.000139  & 0.000121  \\
6 & 6 & 11 & 0.002995  & 0.002671  & -0.000150 & -0.000183 \\
6 & 6 & 13 & 0.002651  & 0.002331  & -0.000029 & -0.000072 \\
\hline
\end{tabular}
\captionsetup{justification=Justified}
\caption{{The OBDME values for $N_{\rm{out}} = 4,5,6$, for $J=0$ and $\tau=0$ (columns 4 and 5) and 1 (columns 6 and 7). Here, $\beta$ ($\alpha$) represents the incoming (outgoing) state, with $N$ being the  harmonic oscillator shell number. The valence space for the shell model calculations does not extend to these states.}
\label{tab:OBDM6Shell}}

\end{table}

\pagebreak
\section{Form factors \label{appendix:FF}}

We note that as $F^{(p,n)}_{X,Y} (q^2)= F^{(n,p)}_{X,Y} (q^2)$ for $X=Y$, we omit repetitive expressions. 

\subsection{Form factors from unprojected states}
\noindent $
    F_M^{(p,p)}(\text{MF}) = e^{-2 y} (323.998 - 635.381 y + 425.649 y^2 - 109.824 y^3 + 
   8.05299 y^4 + 0.248609 y^5 
   + 0.0801673 y^6 - 0.0138916 y^7 + 
   0.000244998 y^8 + 2.64968\times10^{-6} y^9 + 3.53828\times10^{-6} y^{10} 
   - 2.61022\times10^{-7} y^{11} + 5.14035\times10^{-9} y^{12})
$ \\

\noindent $ F_M^{(p,n)}(\rm{MF}) = e^{-2 y} (395.989 - 821.937 y + 600.211 y^2 - 182.005 y^3 + 
   20.8396 y^4 - 0.450053 y^5 
    + 0.0725047 y^6 - 0.0174471 y^7 + 
   0.000884679 y^8 - 0.0000176747 y^9 + 2.5208\times10^{-6} y^{10}
   - 1.82398\times10^{-7} y^{11} + 3.63556\times10^{-9} y^{12})$ \\

\noindent$F_M^{(n,n)}(\rm{MF}) = e^{-2 y} (483.975 - 1060.03 y + 837.688 y^2 - 290.781 y^3 + 
   44.1949 y^4 - 2.56482 y^5 
   + 0.130221 y^6 - 0.0180212 y^7 + 
   0.00111469 y^8 - 0.0000262626 y^9 + 1.79677\times10^{-6} y^{10} 
   - 
   1.27439\times10^{-7} y^{11} + 2.57128\times10^{-9} y^{12})$

\vspace{5mm}

\noindent $F_{\Phi''}^{(p,p)}(\rm{MF}) = e^{-2 y} (0.434581 - 0.273583 y + 0.0680469 y^2 - 0.0120504 y^3 - 
   0.0000907495 y^4 
   + 0.000542141 y^5 - 0.0000741677 y^6 + 0.0000115616 y^7 
   + 1.28506\times10^{-6} y^8 - 2.15953\times10^{-7} y^9 + 6.49007\times10^{-9} y^{10})$\\

\noindent $F_{\Phi''}^{(p,n)}(\rm{MF}) = e^{-2y} (0.63244 - 0.626383 y + 0.215766 y^2 - 0.0347006 y^3 + 
   0.0020224 y^4 + 0.000810892 y^5 - 0.000202625 y^6 + 
   9.42963\times10^{-6} y^7 + 7.69811\times10^{-7} y^8 - 1.01994\times10^{-7} y^9 + 
   3.11955\times10^{-9} y^{10})$\\

\noindent $F_{\Phi''}^{(n,n)}(\rm{MF}) = e^{-2y} (0.920381 - 1.24372 y + 0.60376 y^2 - 0.122635 y^3 + 
   0.00700865 y^4 + 0.00102126 y^5 - 0.000168656 y^6 + 
   6.49523\times10^{-6} y^7 + 4.43646\times10^{-7} y^8 - 4.81567\times10^{-8} y^9 + 
   1.49946\times10^{-9} y^{10})$\\

\vspace{5mm}

\noindent $F_{M\Phi''}^{(p,p)}(\rm{MF}) = e^{-2 y} (-11.8661 + 15.3701 y - 6.09368 y^2 + 1.01117 y^3 - 
   0.0809302 y^4 - 0.0146219 y^5 + 0.00540176 y^6 - 0.000122993 y^7 - 
   9.72973\times10^{-6} y^8 - 2.33849\times10^{-6} y^9 + 2.42743\times10^{-7} y^{10} - 
   5.77592\times10^{-9} y^{11})$ \\

\noindent $F_{M\Phi''}^{(p,n)}(\rm{MF}) = e^{-2y} (-17.2685 + 28.5999 y - 16.2046 y^2 + 3.6749 y^3 - 
   0.242945 y^4 - 0.0168178 y^5 + 0.00121347 y^6 + 0.000132951 y^7 - 
   6.80239\times10^{-6} y^8 - 1.24534\times10^{-6} y^9 + 1.1507\times10^{-7} y^{10} - 
   2.77628\times10^{-9} y^{11})$\\

\noindent $F_{M\Phi''}^{(n,p)}(\rm{MF}) = e^{-2y} (-14.5026 + 20.4471 y - 9.27058 y^2 + 1.87534 y^3 - 
   0.201909 y^4 - 0.0103326 y^5 + 0.00878963 y^6 - 0.000705658 y^7 + 
   0.0000145042 y^8 - 1.69625\times10^{-6} y^9 + 1.69197\times10^{-7} y^{10} - 
   4.08507\times10^{-9} y^{11})$ \\
   
\noindent $F_{M\Phi''}^{(n,n)}(\rm{MF}) = e^{-2y} (-21.1055 + 37.3732 y - 23.3309 y^2 + 6.27645 y^3 - 
   0.663488 y^4 + 0.00214968 y^5 + 0.00365141 y^6 - 0.000202823 y^7 + 
   5.37524\times10^{-6} y^8 - 9.01819\times10^{-7} y^9 + 8.01898\times10^{-8} y^{10} - 
   1.96355\times10^{-9} y^{11})$ \\

\subsection{Form factors from particle number projected states}

\noindent $F_{M}^{(p,p)}(\rm{MF-P}) = e^{-2 y} (324.9998 - 634.172 y + 422.952 y^2 - 107.706 y^3 + 
   7.32564 y^4 + 0.371389 y^5 + 0.0686065 y^6 - 0.0130912 y^7 + 
   0.000188002 y^8 + 9.05031\times10^{-6} y^9 + 2.9477\times10^{-6} y^{10} - 
   2.33863\times10^{-7} y^{11} + 4.67192\times10^{-9} y^{12})$ \\

\noindent $F_{M}^{(p,n)}(\rm{MF-P}) = e^{-2y} (395.994 - 820.429 y + 596.921 y^2 - 179.392 y^3 + 
   19.8907 y^4 - 0.275749 y^5 + 0.0539233 y^6 - 0.0159422 y^7 + 
   0.000790937 y^8 - 0.0000117047 y^9 + 2.03435\times10^{-6} y^{10} - 
   1.57403\times10^{-7} y^{11} + 3.15214\times10^{-9} y^{12})$   \\

\noindent $F_{M}^{(n,n)}(\rm{MF-P}) = e^{-2 y} (483.985 - 1058.15 y + 833.666 y^2 - 287.584 y^3 + 
   42.9972 y^4 - 2.33552 y^5 + 0.105144 y^6 - 0.0158995 y^7 + 
   0.000979656 y^8 - 0.0000199051 y^9 + 1.40334\times10^{-6} y^{10} - 
   1.05941\times10^{-7} y^{11} + 2.12675\times10^{-9} y^{12})$\\ \\ \\ \\

\noindent $F_{\Phi''}^{(p,p)}(\rm{MF-P}) = e^{-2 y} (0.408643 - 0.252629 y + 0.0637581 y^2 - 0.011918 y^3 - 
   0.0000317666 y^4 + 0.000508242 y^5 - 0.0000730376 y^6 + 
   0.0000121761 y^7 + 1.28512\times10^{-6} y^8 - 2.18788\times10^{-7} y^9 + 
   6.55054\times10^{-9} y^{10})$ \\

\noindent $F_{\Phi''}^{(p,n)}(\rm{MF-P}) = e^{-2 y}(0.603176 - 0.592923 y + 0.20307 y^2 - 0.0330578 y^3 + 
  0.00202836 y^4 + 0.000795586 y^5 - 0.000199969 y^6 + 
  8.89218\times10^{-6} y^7 + 8.07867\times10^{-7} y^8 - 1.02983\times10^{-7} y^9 + 
  3.11902\times10^{-9} y^{10})$   \\

\noindent $F_{\Phi''}^{(n,n)}(\rm{MF-P}) = e^{-2 y}(0.890315 - 1.19996 y + 0.579046 y^2 - 0.11572 y^3 + 0.00602047 y^4 + 
  0.00107123 y^5 - 0.000164302 y^6 + 5.78596\times10^{-6} y^7 + 
  4.78185\times10^{-7} y^8 - 4.84678\times10^{-8} y^9 + 1.48511\times10^{-9} y^{10})$  \\

\vspace{5mm}

\noindent $F_{M\Phi''}^{(p,p)}(\rm{MF-P}) = e^{-2 y}(-11.5065 + 14.8178 y - 5.8288 y^2 + 0.974184 y^3 - 0.0813589 y^4 - 
  0.0150261 y^5 + 0.00550428 y^6 - 0.000111431 y^7 - 
  0.0000125089 y^8 - 2.09598\times10^{-6} y^9 + 2.30844\times10^{-7} y^{10} - 
  5.53205\times10^{-9} y^{11})$ \\

\noindent $F_{M\Phi''}^{(p,n)}(\rm{MF-P}) = e^{-2 y}(-16.9842 + 28.0673 y - 15.82 y^2 + 3.53497 y^3 - 0.215039 y^4 - 
  0.0199046 y^5 + 0.00140454 y^6 + 0.000125178 y^7 - 
  7.0951\times10^{-6} y^8 - 1.15511\times10^{-6} y^9 + 1.08909\times10^{-7} y^{10} - 
  2.63407\times10^{-9} y^{11})$ \\

\noindent $F_{M\Phi''}^{(n,p)}(\rm{MF-P}) = e^{-2 y}(-14.0633 + 19.7206 y - 8.88652 y^2 + 1.8085 y^3 - 0.201656 y^4 - 
  0.010219 y^5 + 0.00886869 y^6 - 0.000696287 y^7 + 
  0.0000121262 y^8 - 1.47187\times10^{-6} y^9 + 1.55296\times10^{-7} y^{10} - 
  3.73247\times10^{-9} y^{11})$ \\

\noindent $F_{M\Phi''}^{(n,b)}(\rm{MF-P}) = e^{-2 y}(-20.7581 + 36.6808 y - 22.8038 y^2 + 6.0748 y^3 - 0.621238 y^4 - 
  0.00265604 y^5 + 0.00392472 y^6 - 0.000210361 y^7 + 
  4.97847\times10^{-6} y^8 - 8.06913\times10^{-7} y^9 + 7.32646\times10^{-8} y^{10} - 
  1.7772\times10^{-9} y^{11})$

\end{widetext}
\pagebreak

\bibliography{references.bib}

\end{document}